\documentclass[a4paper,7pt]{article} 
\usepackage[letterpaper,left=1cm,right=1cm,top=2.5cm,bottom=4cm]{geometry}
\usepackage[english,activeacute]{babel}
\usepackage[utf8]{inputenc} 
\usepackage{setspace}
\usepackage{graphics}
\usepackage[demo]{graphicx}

\DeclareMathAlphabet{\mathpzc}{OT1}{pzc}{m}{it} 
\usepackage[usenames]{color}
\usepackage{hyperref}
\usepackage{longtable}
\usepackage{amsmath}
\usepackage{epigraph}
\usepackage{indentfirst}
\usepackage{verbatim}
\usepackage{array}
\usepackage{amssymb}
\usepackage{subfigure}
\usepackage{fancyhdr}
\usepackage{amsmath}
\usepackage{mathtools}
\usepackage{lipsum}
\usepackage{ulem}
\usepackage{cuted}
\usepackage{mathrsfs}
\usepackage[small,bf]{caption}
\usepackage[numbers]{natbib}
\usepackage{cancel}
\usepackage{braket} 
\usepackage{epstopdf}
\usepackage{tikz-feynman}
\usepackage{feynmp}
\usepackage{authblk}
\usepackage{tikz-feynman}
\tikzfeynmanset{compat=1.1.0}
\usepackage{tikz}
\usetikzlibrary{trees}
\usetikzlibrary{decorations.pathmorphing}
\usetikzlibrary{decorations.markings}
\usepackage[titletoc,toc,page]{appendix}
\def\bb{\ensuremath\bold}
\usepackage{multicol}
 \title{Path integral description and direct interaction approximation for elastic plate turbulence}
\author[1]{Ignacio Pavez}
\author[1]{Gustavo Düring}
\affil[1]{Instituto de F\'{i}sica, Pontificia Universidad Cat\'{o}lica de Chile, Casilla 306, Santiago, Chile}
\begin{document}
\maketitle

\begin{abstract}
In this work, we apply the Martin-Siggia-Rose path integral formalism to the equations of a thin elastic plate. Using a diagrammatic technique, we obtain the direct interaction approximation (DIA) equations to describe the evolutions of the correlation function and the response function of the fields. Consistent with previous results, we show that DIA equations for elastic plates can be derived from a non-markovian stochastic process and that in the weakly nonlinear limit, the DIA equations lead to the kinetic equation of wave turbulence theory. We expect that this approach will allow a better understanding of the statistical properties of wave turbulence and that DIA equations can open new avenues for understanding the breakdown of weakly nonlinear turbulence for elastic plates. 
\end{abstract}

\begin{multicols}{2}
\section{Introduction}
Turbulence represents one of the long-standing and unsolved problems which has challenged physicists for centuries. 
Although the turbulence phenomenon is rooted in the dynamics of fluid flows, turbulence-like behavior is rather ubiquitous in nature. 
A large class of systems ranging from quantum scales to astrophysical ones \cite{saur2002evidence,wilson2013experimental,denissenko2007gravity}  are characterized by out-of-equilibrium randomly fluctuating waves, a state referred to as wave turbulence (WT). These systems have many similarities with hydrodynamic turbulence. They display a Richardson-like cascade with an energy flux from large scales toward small scales, leading to a stationary out-of-equilibrium solution for the energy spectrum, called Kolmogorov-Zakharov\cite{zakharov1975statistical,zakharov2012kolmogorov}. The significant advantage of random wave systems is that the dynamics can be considered typically as weakly nonlinear, in stark opposition to hydrodynamic turbulence. In consequence, it is possible to establish a closed kinetic equation for wave amplitudes evolution which capture the WT behavior \cite{peierls1929kinetischen,benney1966nonlinear,benney1969random,hasselmann1962non}. The wave turbulence theory has been extensively applied to quantum turbulence \cite{vinen2002quantum,kozik2005scale},  surface water waves\cite{annenkov2001numerical,falcon2007observation,bedard2013non}, Alfvén waves \cite{ng1997scaling} and even to describe the preheating stage of inflation in the early Universe \cite{micha2003relativistic}. 

The case of elastic plates \cite{during2006weak} has been considered experimentally and numerically in several works \cite{boudaoud2000dynamics,chopin2008liquid,mordant2008there,mordant2010fourier,cobelli2009space,during2017wave,during2018exact,hassaini2019elastic,yokoyama2013weak,yokoyama2014identification}. It has been shown to be an excellent playground for WT due to the simplicity of the experimental setup and of the dynamic equations for the surface elevation. Although important progress has been made, a full understanding of the probability density function evolution, as well as the nonlinear response of elastic plates, remains unclear. In addition, in the past decade, it has been reported that the weakly nonlinear regime breaks down at increasing forcing, and a new regime emerges \cite{miquel2013transition}\cite{yokoyama2013weak}. Under strongly nonlinear effects, elastic plates display an energy spectrum different from the WT theory prediction. Intermittency, which is responsible for breaking self-similarity, is also observed \cite{during2019strong}. Such a phenomenon, which is well known in hydrodynamic turbulence \cite{frisch1995turbulence}, goes beyond the standard description of the WT theory and remains largely unclear. In this work, using the Martin-Siggia-Rose formalism\cite{martin1973statistical,gauthier1981testing,castellani2005spin} we construct a path integral description for the stochastic dynamics of plates. This statistical description enables a full representation of the probability distribution function and a clear description of correlations and responses of the system under external perturbation. We expect that this formalism will provide new insights into the statistical properties of weakly nonlinear elastic plates and might elucidate some properties beyond the breakdown of the wave turbulence theory.

 Different non-perturbative approximations have been developed to understand the properties of strong turbulence, ranging from the introduction of the so-called eddy viscosity constant \cite{boussinesq1877essai} to methods recycled from quantum field theory.  Theories based on renormalization have led to several descriptions such as Wyld's analysis \cite{wyld1961formulation}, Local Energy Transfer Theory (LET) \cite{edwards1971local,mccomb1974local}, Direct Interaction Approximation (DIA)\cite{kraichnan1959structure} and others \cite{mccomb1990physics}. Each of them leads to different results and descriptions, showing some degree of similarity  \cite{mccomb1984numerical,mccomb1990physics}, but in general, fail to give an accurate description of hydrodynamic turbulence. DIA presents several advantages compared to other renormalization methods. First, it lays on an exact correspondence with an underlying stochastic system which guarantees a well-defined evolution. Second, it correctly yields to the kinetic equation when applied to weakly nonlinear wave systems \cite{krommes2002fundamental}. Here, using the path integral formalism, we derive the DIA equations for elastic plates to extend previous results to cubic nonlinear systems which are not phase invariant.  We obtain an extra non-trivial term in the DIA arising from one-loop correction diagrams, which could impact the phase mixing at increasing wave amplitude. This extra term does not lead to energy transfer along modes and in the weakly nonlinear regime, it is responsible for the frequency renormalization \cite{newell2001wave,zakharov2012kolmogorov,during2009symmetry,during2006weak,galtier2017turbulence}.

DIA equations have been largely applied to different nonlinear systems from the  Duffing Oscillator \cite{morton1970consolidated,gauthier1981testing} to Langmuir Turbulence in plasmas or the nonlinear Schr\"odinger equation (NLS) \cite{dubois1981statistical,sun1985statistical}.   For strongly nonlinear regimes, DIA is known to yield relatively unsatisfactory results \cite{kraichnan1959structure,sun1985statistical,sun1986statistical,orszag1972numerical}. However, when a modified version of the DIA is imposed, the prediction compares well with numerical results \cite{frederiksen2004regularized,krommes2002fundamental,bowman1993realizable}. Thus, this seems to be the best starting point to study the breakdown of the WT theory, due to its simplicity and realizability. 

\section{Föppl-von Kármán equations for plates}
Thin elastic plates are modeled by the Föppl-von Kármán(FvK) equations \cite{landau1986theory}. These equations describe the time evolution of the vertical displacement $\zeta$ and the Airy stress function $\chi$ where
 \begin{eqnarray}
 \label{FvK1}
 \rho\frac{\partial^2\zeta}{\partial t^2}(\bb{x},t)&=& -\frac{Eh^2}{12(1-\sigma^2)}\Delta^2\zeta(\bb{x},t)+\{\zeta,\chi\}(\bb{x},t),\\
 \label{FvK2}
 \frac{1}{E}\Delta^2\chi(\bb{x},t)&=&-\frac{1}{2}\{\zeta,\zeta\}(\bb{x},t)
 \end{eqnarray}
and $\{f,g\}:=f_{xx}g_{yy}+f_{yy}g_{xx}-2f_{xy}g_{xy}$. The physical parameters are given by the mass density $\rho$, the thickness $h$, the Young's modulus $E$ and the Poisson ratio $\sigma$. The linear regime is characterized by bending waves, which display a relation dispersion $\omega_{\mathbf{k}}= \sqrt{\frac{Eh^2}{12(1-\sigma^2)\rho}}|\bb{k}|^2$. Using the canonical  change of variables $A_{\mathbf{k}}(t):=\frac{1}{\sqrt{2}}\left(\sqrt{\omega_{\mathbf{k}}\rho h}\zeta_{\mathbf{k}}(t)+\frac{i}{\sqrt{\omega_{\mathbf{k}}\rho h}}p_{\mathbf{k}}(t)\right)$   for the Fourier transform of the displacement field $\zeta$ and for the Fourier transforms of the momentum $p=\rho\partial_t \zeta$, the FvK equations transform to the general form of cubic nonlinear wave systems \cite{during2006weak}  
\begin{small}
\begin{eqnarray}
\label{ecndin}
  \left(\frac{\partial}{\partial t}+is\omega_{\textbf{k}}\right)A^s_{\textbf{k}}(t)& =&\sum_{s_1s_2s_3}\int L^{ss_1s_2s_3}_{\textbf{k}\textbf{k}_1\textbf{k}_2\textbf{k}_3} A^{s_1}_{\textbf{k}_1}(t)A^{s_2}_{\textbf{k}_2}(t)A^{s_3}_{\textbf{k}_3}(t)\nonumber\\
  &&\times\delta(\textbf{k}_1+\textbf{k}_2+\textbf{k}_3-\textbf{k})d\textbf{k}_{123}.
\end{eqnarray}
\end{small}
We utilize Newell's notation \cite{newell2001wave} : $A^{+}_{\mathbf{k}}(t):=A_{\mathbf{k}}(t)$, $A^{-}_{\mathbf{k}}(t):=A^{*}_{-\mathbf{k}}(t)$ and $d\textbf{k}_{123}:=d\textbf{k}_{1}d\textbf{k}_{2}d\textbf{k}_{3}$.  It should be noted that for finite size plates, integrals in Fourier space must be replaced with sums, but the entire formalism developed in the following remains unaltered.
For the elastic plate:
\begin{eqnarray}
L^{ss_1s_2s_3}_{\mathbf{k}\mathbf{k}_1\mathbf{k}_2\mathbf{k}_3}&=& \frac{-is}{6\pi^2}X_{-\mathbf{k}}X_{\mathbf{k}_1}X_{\mathbf{k}_2}X_{\mathbf{k}_3}\\
&&\times \left(T_{-\mathbf{k}\mathbf{k}_1\mathbf{k}_2\mathbf{k}_3}+T_{-\mathbf{k}\mathbf{k}_2\mathbf{k}_1\mathbf{k}_3}+T_{-\mathbf{k}\mathbf{k}_3\mathbf{k}_2\mathbf{k}_1}\right)\nonumber
\end{eqnarray}
with 
\begin{eqnarray}
T_{\mathbf{k_1}\mathbf{k_2}\mathbf{k_3}\mathbf{k_4}}=\frac{1}{16}\left( \frac{1}{|\mathbf{k_1}+\mathbf{k_2}|^2}+\frac{1}{|\mathbf{k_3}+\mathbf{k_4}|^2}\right)(\mathbf{k_1}\times\mathbf{k_2})^2(\mathbf{k_3}\times\mathbf{k_4})^2\nonumber.
\end{eqnarray}

The canonical equations (\ref{ecndin}) can be encountered in several  nonlinear wave systems such as gravity waves \cite{dyachenko2004weak,zakharov1966energy} or capillary waves \cite{zakharov1967weak,during2009symmetry}; however, beyond the weakly nonlinear regime, higher order corrections need to be included for those systems.  Another important example is the Gross-Pitaevskii  \cite{nazarenko2006wave,dyachenko1992optical,nazarenko2007freely,zakharov1985hamiltonian}  or NLS equations, which have been extensively studied  including a description using DIA \cite{dubois1981statistical,hansen1981cubic}. However, it is known that the phase invariance of NLS implies an extra conserved quantity, leading to a complex statistical description even for initially weakly non linear amplitudes  \cite{dyachenko1992optical}. It is also interesting to observe that ideal incompressible fluids can be recast in a similar form making use of  Clebsch pair $\mu$ and $\lambda$, where the velocity field reads $\mathbf{v}=-\frac{1}{\rho} \Delta^{-1}\nabla \times (\nabla \lambda \times \nabla \mu)$ with $\rho$ the density. The dynamic equation in Fourier space thus takes the form (\ref{ecndin}) with $\omega_k=0$,  $A_\mathbf{k}=(\mu_\mathbf{k}+i \lambda_\mathbf{k})/\sqrt{2}$ and the scattering matrix $L^{s s_1 s_2 s_3}_{{\mathbf k} {\mathbf k}_1{\mathbf k}_2{\mathbf k}_3}$ given in \cite{zakharov2012kolmogorov}.

Canonical equation (\ref{ecndin}) can be shown to derive from a Hamiltonian structure $$is\partial_t A^s_{\mathbf{k}}(t)=\frac{\delta H}{\delta A^{-s}_{\mathbf{k}}(t)}$$ where for the elastic plate
\begin{small}
\begin{eqnarray}
H&=&\int d\bb{k} \omega_{\bb{k}}A_{\bb{k}}(t)A^{*}_{\bb{k}}(t)+\frac{1}{4(2\pi)^2}\int d\bb{k}_{1234}X_{\bb{k}_1}X_{\bb{k}_2}X_{\bb{k}_3}X_{\bb{k}_4}T_{\bb{k_1k_2};\bb{k_3k_4}}\nonumber\\
&&\times \sum_{s_1s_2s_3s_4}A^{s_1}_{\bb{k}_1}(t)A^{s_2}_{\bb{k}_2}(t)A^{s_3}_{\bb{k}_3}(t)A^{s_4}_{\bb{k_4}}\delta(\bb{k_1+k_2+k_3+k_4}).
\end{eqnarray}
\end{small}
To obtain an out-of-equilibrium turbulent state, we have to include forcing and damping terms into the dynamics
\begin{equation}
\label{generalwave}
is\partial_t A^s_{\mathbf{k}}(t)=\eta^s_{\mathbf{k}}-is\gamma_{\mathbf{k}}A^s_{\mathbf{k}}(t)+\frac{\delta H}{\delta A^{-s}_{\mathbf{k}}(t)},
\end{equation}
where $\gamma_{\bf{k}}$ is the damping term and  $\eta^s_{\textbf{k}}$ the external force. We will assume $\eta^s_{\textbf{k}}$ to be a complex random variable with a Gaussian distribution and delta correlated ,i.e, $\langle \eta^{s}_{\textbf{k}}(t)\eta^{z}_{\textbf{q}}(t')\rangle = 2F_{\textbf{q}}\delta(t-t')\delta_{-s,z}\delta(\textbf{k+q})$ with $(\eta^{s_1}_{-\mathbf{k}_1}(t_1))^*=\eta^{-s_1}_{\mathbf{k}_1}(t_1)$  \footnote{The star superscript denotes the complex conjugate.} .  To study the statistical properties of this system, we will write in the next chapter the path integral description following the standard Martin-Siggia-Rose formalism \cite{martin1973statistical,gauthier1981testing,castellani2005spin}.

\section{Path Integral formalism}

The key idea in constructing the path integral functional is to realize that  each solution $A^{s}_{\bf{k}}$ depends on the specific realization of the noise $\eta^s_{\bf{k}}$. Thus, we can obtain a probability distribution function (PDF) for the field $A^{s}_{\bf{k}}$ combining the PDF of the gaussian noise and the constraint of the function $A^{s}_{\bf{k}}$ to the stochastic differential equation (SDE).  The average for an observable $O=O(A^{s_1}_{k_1}(t),\ldots,A^{s_n}_{k_n}(t))$ is
\begin{small}
 \begin{eqnarray}
 \label{ZNL}
 \langle O \rangle &=& \int \mathcal{D}\eta^s_k(t) P(\eta^s_{k}(t))O(A^{s_1}_{k_1,\eta^{s_1}_{k_1}},\ldots, A^{s_n}_{k_n,\eta^{s_n}_{k_n}}),
 \end{eqnarray}
 \end{small}
where $\mathcal{D}\eta^s_k(t):= \lim\limits_{N\to \infty}\prod^{k_n}_{k=k_1} \prod^{s_n}_{s=s_1}\prod^{N}_{j=1}d\eta^s_{k,j}$( the subindex $j$ is for the time discretization),  $P(\eta^s_{k}(t))$ is a Gaussian PDF and  $A^{s_n}_{k_n,\eta^{s_n}_{k_n}}$ is the field under a given realization of the noise. Inserting a Dirac delta function with the SDE as an argument allows us to rewrite the average as integrating over the field $A^s_{\bf{k}}$. 
Averaging over the noise, by making use of the integral representation of the delta function leads to
\begin{small}
\begin{eqnarray}
\label{Z}
 \langle O \rangle&=& \int \mathcal{D}A^s_{\mathbf{k}}(t) \mathcal{D}\tilde{A}^s_{\mathbf{k}}(t) O(A^{s}_{k}(t),\ldots)e^{-\mathcal{S}[A^{s}_{\mathbf{k}}(t),\tilde{A}^{s}_{\mathbf{k}}(t)]}
 \end{eqnarray}
\end{small}
where $\tilde{A}^s_{\mathbf{k}} $ corresponds to the integration variable from the Dirac delta integral representation and the MSR action reads as
\begin{small}
\begin{eqnarray}
&&\mathcal{S}[A^{s}_{\mathbf{k}}(t),\tilde{A}^{s}_{\mathbf{k}}(t)]\coloneqq\nonumber\\
&&\sum_{s}\int d\mathbf{k} dt \tilde{A}^{s}_{\mathbf{k}}(t)\big(\partial_t A^{s}_{\mathbf{k}}(t)+\gamma_{\mathbf{k}}A^s_{\mathbf{k}}(t)+i s \frac{\delta H}{\delta A^{-s}_{\mathbf{k}}(t)}-F_{\mathbf{k}}\tilde{A}^{-s}_{-\mathbf{k}}(t)\big).\nonumber\\
\end{eqnarray}
\end{small}
Therefore, we can define the generating functional  in terms of the auxiliary fields $J^s_{\mathbf{k}}(t)$ and $\tilde{J}^s_{\mathbf{k}}(t)$ as
\begin{small}
\begin{equation}
\label{ZJJ}
Z[J,\tilde{J}]=\int \mathcal{D}A^s_{\mathbf{k}} \mathcal{D}\tilde{A}^s_{\mathbf{k}}e^{-\mathcal{S}[A^{s}_{\mathbf{k}}(t),\tilde{A}^{s}_{\mathbf{k}}(t)]+\sum_s\int d\mathbf{k} \int \tilde{J}^s_{\bf{k}}(t)A^{s}_{\bf{k}}(t)+J^s_{\bf{k}}(t)\tilde{A}^{s}_{\bf{k}}(t)dt},
\end{equation}
\end{small}
where $Z[0,0]=1$. Correlation functions can then be easily obtained by differentiating over the field $\tilde{J}$ and $J$. In particular, the two point correlation function $C^{l_1l_2}_{\mathbf{p_1p_2}}(t_1,t_2):= \big\langle A^{l_1}_{\mathbf{p_1}}(t_1)A^{l_2}_{\mathbf{p_2}}(t_2)\big\rangle$ is given by 
\begin{eqnarray}
\label{CderivZ}
C^{l_1l_2}_{\mathbf{p_1p_2}}(t_1,t_2)&=&\left.\frac{\delta^2 Z[J,\tilde{J}]}{\delta \tilde{J}^{l_2}_{\bf{p}_2}(t_2)\delta \tilde{J}^{l_1}_{\bf{p}_1}(t_1)}\right|_{J=\tilde{J}=0}.
\end{eqnarray}
The response function defined as $R^{l_1-l_2}_{\bf{p_1}\bf{-p_2}}(t_1,t_2):=\left\langle\frac{\delta A^{l_1}_{\bf{p_1}}(t_1)}{\delta \eta^{l_2}_{\bf{p_2}}(t_2)}\right\rangle$  can also be obtained directly from the generating functional. This quantity simply measures (in average) the infinitesimal change of the field with respect to a source. It can be shown (Appendix \ref{D}) that 
\begin{eqnarray}
\label{RderivZ}
R^{l_1-l_2}_{\mathbf{p_1-p_2}}(t_1,t_2)&=&\left.\frac{\delta^2 Z[J,\tilde{J}]}{\delta J^{l_2}_{\bf{p}_2}(t_2)\delta \tilde{J}^{l_1}_{\bf{p}_1}(t_1)}\right|_{J=\tilde{J}=0}.
\end{eqnarray}

Performing the path integral for the generating functional (\ref{ZJJ})  is extremely complicated and only in the absence of the nonlinear term, called the \textit{free} case, a general solution can be found. The moment generating functional for the \textit{free} case $Z_0[J,\tilde{J}]$ is obtained by keeping the linear term of the Hamiltonian $H\rightarrow H_0=\int d\bb{k} \omega_{\bb{k}}A_{\bb{k}}(t)A^{*}_{\bb{k}}(t)$. The \textit{free} generating functional then reads as (see details in appendix \ref{A})
\begin{eqnarray}
\label{Z_0}
Z_0[J,\tilde{J}]= e^{-\frac{1}{2}\sum_{s_1s_2}\int d\mathbf{k_1}d\mathbf{k_2}\int dtdt' j^{s_1,T}_{\mathbf{k_1}}(t)G^{s_1s_2}_{\bb{k_1k_2}}(t,t')j^{s_2}_{\bb{k_2}}(t')}
\end{eqnarray}
where 
\begin{small}
\begin{equation}
G^{s_1s_2}_{\bb{k_1k_2}}(t,t')=
\begin{pmatrix}
C^{s_1s_2}_{0,\bb{k_1k_2}}(t,t') & R^{s_1-s_2}_{0,\bb{k_1-k_2}}(t,t')\\
R^{s_2-s_1}_{0,\bb{k_2-k_1}}(t',t) & 0
\end{pmatrix}
\end{equation}
\end{small}
with $j^{s_1}_{\mathbf{k_1}}(t)\coloneqq \begin{pmatrix} \tilde{J}^{s_1}_{\mathbf{k_1}}(t) \\ J^{s_1}_{\mathbf{k_1}}(t)\end{pmatrix}$ and the superscript $T$ denotes transpose. The subindex 0 hereafter refers to functions associated with the linear Hamiltonian $H_0$. It is to be noted that we express this functional only in terms of the correlation and response function; this is an outstanding advantage of this construction and the reason will be clarified later.
The  explicit solutions for the correlation and response functions are known for the \textit{free} case. Considering the dynamic equation (\ref{generalwave}) with the linear Hamiltonian $H_0$, one can easily obtain closed equations for the two-point correlation and response function
 \begin{eqnarray}
 \left(\partial_t+is\Omega^s_{\mathbf{k}}\right)C^{sl_2}_{0,\mathbf{k}\mathbf{p}_2}(t,t_2)&=&2F_{\mathbf{p}_2}R^{l_2s}_{0,\mathbf{p}_2\mathbf{k}}(t_2,t) \nonumber\\
  \left(\partial_t+is\Omega^s_{\mathbf{k}}\right)R^{s-l_2}_{0,\mathbf{k}-\mathbf{p}_2}(t,t_2)&=&\delta_{sl_2}\delta(\mathbf{k}-\mathbf{p}_2)\delta(t-t_2)
 \end{eqnarray}
where $\Omega^{s}_{\bb{k}}:=\omega_{\bb{k}}-is\gamma_{\bb{k}}$ and we made use of the identity  $R^{s-l_2}_{0,\mathbf{k}-\mathbf{p}_2}(t,t_2)= \frac{1}{2F_{\bb{-p_2}}}\left\langle A^{s}_{\bb{k}}(t)\eta^{-l_2}_{\bb{-p_2}}(t_2)\right\rangle$(see Appendix \ref{D} for details). Finally the zero order correlation and response  function are given by:
\begin{small}
\begin{eqnarray}
\label{R0}
R^{l_1-l_2}_{\bb{p_1-p_2}}(t_1,t_2)&=&\delta_{l_1l_2}\delta(\bb{p_1-p_2})e^{-il_1\Omega^{l_1}_{\bb{p_1}}(t_1-t_2)}\nonumber\\
\label{C0}
C^{l_1l_2}_{0,\bb{p_1p_2}}(t_1,t_2)&=&C^{l_1l_2}_{0,\bb{p_1p_2}}(0,0)e^{-i(l_1\Omega_{\bb{p_1}}t_1+l_2\Omega_{\bb{p_2}}t_2)}\nonumber\\
&&+2F_{\bb{p_2}}\delta_{l_1,-l_2}e^{il_1\omega_{\bb{p}_2}(t_2-t_1)}t_1\theta(t_2-t_1)\nonumber\\
&&+2F_{\bb{p_2}}\delta_{l_1,-l_2}e^{il_1\omega_{\bb{p}_2}(t_2-t_1)}t_2\theta(t_1-t_2)\nonumber\\
\end{eqnarray}
\end{small}
where $\theta$ is the step function.

 \section{DIA equations}
 
 Calculating the path integral of the full nonlinear generating functional is a complicated problem, and an analog expression to (\ref{Z_0}) for the exact $Z$ is not known. Thus, to study the nonlinear regime, some level of approximation needs to be used. The Direct interaction approximation (DIA) corresponds to a particular resummation scheme which leads to a closed set of coupled nonlinear integro-differential equations for the correlation and response functions. In contrast to the original derivation, the use of the Martin-Siggia-Rose formalism exposes naturally the role of the response function in the statistical dynamics, as shown in the equation for the free generating functional (\ref{Z_0}).

The starting point is to consider perturbatively the non linear term of the generating functional (\ref{ZJJ}) and  expand the exponential into a power series. Every term in the series will corresponds to functional derivatives of $Z_0$ with respect to the auxiliar fields. Thus, we can write: $Z= Z_0[J,\tilde{J}]+Z_1[J,\tilde{J}]+Z_2[J,\tilde{J}]+\ldots$. Consequently, the correlation $C^{l_1l_2}_{\mathbf{p_1p_2}}(t_1,t_2)$ and response function $R^{l_1-l_2}_{\mathbf{p_1-p_2}}(t_1,t_2)$ can be written as  expansions series making use of  (\ref{CderivZ}) and (\ref{RderivZ}), where the zero order term correspond precisely to the free correlation $C^{l_1l_2}_{0,\mathbf{p_1p_2}}(t_1,t_2)$ and the free response $R^{l_1-l_2}_{0,\mathbf{p_1-p_2}}(t_1,t_2)$, respectively (see details in Appendix \ref{B}).
To obtain an intuitive understanding of the procedure, we use a diagram technique.
So if we name every function ( independent of the indices):\\
$C \rightarrow $
\begin{tikzpicture}
\begin{feynman}[inline=(a)]
\vertex(b);
\vertex[right=2cm of b] (d);
\diagram*{
    (b)-- [gluon, edge label=\(t_2\hspace{10mm}t_1\)] (d),
};
\end{feynman}
\end{tikzpicture}
\\$C_{0} \rightarrow $
\begin{tikzpicture}
\begin{feynman}[inline=(a)]
\vertex(b);
\vertex[right=2cm of b] (d);
\diagram*{
    (b)-- [scalar, edge label=\(t_2\hspace{10mm}t_1\)] (d),
};
\end{feynman}
\end{tikzpicture}
\vspace{4mm}\\
$R\rightarrow$
\begin{tikzpicture}
\begin{feynman}[inline=(a)]
\vertex(b);
\vertex[right=2cm of b] (d);
\diagram*{
    (b)-- [photon, edge label=\(t_2\hspace{10mm}t_1\)] (d),
};
\end{feynman}
\end{tikzpicture}
\\$R_0\rightarrow $
\begin{tikzpicture}
\begin{feynman}[inline=(a)]
\vertex(b);
\vertex[right=2cm of b] (d);
\diagram*{
    (b)-- [plain, edge label=\(t_2\hspace{10mm}t_1\)] (d),
};
\end{feynman}
\end{tikzpicture}\\
one could find the Feynman diagrams for the correlation and the response functions. However, for practical reasons, we define the differential operator $\hat{\mathcal{L}}^{l_1}_{\mathbf{p_1},t_1}\coloneqq \frac{\partial}{\partial t_1} + il_1\Omega^{l_1}_{\mathbf{p_1}}$ and then expand the quantities $\hat{\mathcal{L}}^{l_1}_{\mathbf{p_1},t_1} C^{l_1l_2}_{\mathbf{p_1p_2}}(t_1,t_2)$  and $\hat{\mathcal{L}}^{l_1}_{\mathbf{p_1},t_1} R^{l_1-l_2}_{\mathbf{p_1-p_2}}(t_1,t_2)$ instead of the pure two-point correlation and response function.  Then the Feynman diagrams up to the second order take the form (showed in Appendix \ref{B}):\\
\end{multicols}
 \begin{eqnarray}
 \label{ExpDiagrams}
\hat{\mathcal{L}}^{l_1}_{\mathbf{p}_1, t_1}
\begin{tikzpicture}
\begin{feynman}[inline=(a)]
\vertex(b);
\vertex[right=2cm of b] (d);
\diagram*{
    (b)-- [gluon, edge label=\(\)] (d),
};
\end{feynman}
\end{tikzpicture}
&=&
\underbrace{\begin{tikzpicture}
\begin{feynman}[inline=(a)]
\vertex(b);
\vertex[right=2cm of b] (d);
\diagram*{
    (b)-- [plain, edge label=\(t_2\hspace{10mm}t_1\)] (d),
};
\end{feynman}
\end{tikzpicture}}_\text{Order R0}+
\underbrace{\begin{tikzpicture}
\begin{feynman}[inline=(e)]
\vertex(a);
\vertex[right=2cm of a] (b);
\vertex[above= of b] (c);
\diagram*{
    (a)-- [scalar, edge label=\(t_2\hspace{10mm}\)] (b),
    (b)-- [scalar, half left, edge label=\(t_1\)] (c),
    (c)-- [scalar, half left, edge label=\(\)] (b),
};
\end{feynman}
\end{tikzpicture}}_\text{Order 1, type A}
+\underbrace{\begin{tikzpicture}
\begin{feynman}[inline=(e)]
\vertex(a);
\vertex[right=2cm of a] (b);
\vertex[above= of b] (c);
\vertex[right=2cm of b] (d);
\diagram*{
    (a)-- [plain, edge label=\(t_2\hspace{10mm}\)] (b),
    (b)-- [scalar, half left, edge label=\(t\)] (c),
    (c)-- [scalar, half left, edge label=\(\)] (b),
    (b)-- [plain, edge label=\(\hspace{10mm}t_1\)] (d)
};
\end{feynman}
\end{tikzpicture}}_\text{Order 1, type RA}
\nonumber\\
&&
+
\underbrace{ \begin{tikzpicture}
\begin{feynman}[inline=(a)]
\vertex(a);
\vertex[right=2cm of a] (b);
\vertex[above= of b] (c);
\vertex[right=2cm of b](d);
\vertex[above= of d] (e);
\diagram*{
    (a)-- [scalar, edge label=\(t_2\hspace{10mm}\)] (b),
    (b)-- [scalar, half left, edge label=\(t\)] (c),
    (c)-- [scalar, half left, edge label=\(\)] (b),
    (b)-- [plain, edge label=\(\hspace{15mm}\)] (d),
    (d)-- [scalar, half left, edge label=\(t_1\)] (e),
    (e)-- [scalar, half left, edge label=\(\)] (d),
};
\end{feynman}
\end{tikzpicture}}_\text{Order 2, type A}
+\underbrace{\begin{tikzpicture}
\begin{feynman}[inline=(a.base)]
\vertex(a);
\vertex[right=2cm of a] (b);
\vertex[above= of b] (c);
\vertex[right=2cm of b] (d);
\vertex[above= of d] (e);
\vertex[right=2cm of d] (g);
\diagram*{
    (a)-- [plain, edge label=\(t_2\hspace{15mm}t'\)] (b),
    (b)-- [scalar, half left, edge label=\(t'\)] (c),
    (c)-- [scalar, half left, edge label=\(\)] (b),
    (b)-- [plain, edge label=\(\hspace{18mm}t\)] (d),
    (d)-- [scalar, half left, edge label=\(t\)] (e),
    (e)-- [scalar, half left, edge label=\(\)] (d),
    (d)-- [plain,edge label=\(\hspace{18mm}t_1\)] (g),
};
\end{feynman}
\end{tikzpicture}}_\text{Order 2, type RA}\nonumber\\
&&+
\underbrace{\begin{tikzpicture}[baseline=(a)]
\begin{feynman}[inline=(a)]
\vertex(a);
\vertex[right=2cm of a] (b);
\vertex[right=2cm of b] (c);
\vertex[above=3cm of b] (e);
\diagram*{
    (a)-- [plain, edge label=\(t_2\hspace{10mm}t\)] (b),
    (b)-- [scalar, half left, edge label=\(\)] (c),
    (b)-- [scalar, edge label=\(\hspace{10mm}t_1\)] (c),
    (b)-- [scalar, half right, edge label=\(\)] (c),
};
\end{feynman}
\end{tikzpicture}}_\text{Order 2, type B}
+
\underbrace{\begin{tikzpicture}[baseline=(a)]
\begin{feynman}[inline=(a)]
\vertex(a);
\vertex[right=1.5cm of a] (b);
\vertex[right=1.5cm of b] (c);
\vertex[above=3cm of b] (e);
\vertex[right=1.5cm of c] (d);
\diagram*{
    (a)-- [plain, edge label=\(t_2\hspace{5mm}t'\)] (b),
    (b)-- [scalar, half left, edge label=\(\)] (c),
    (b)-- [plain, edge label=\(\hspace{5mm}t\)] (c),
    (b)-- [scalar, half right, edge label=\(\)] (c),
    (c)-- [plain,edge label=\(t_1\)](d),
};
\end{feynman}
\end{tikzpicture}}_\text{Order 2, type RB}
\nonumber\\
&&+\underbrace{\begin{tikzpicture}[baseline=(a)]
\begin{feynman}[inline=(a)]
\vertex(a);
\vertex[right=2cm of a] (b);
\vertex[right=2cm of b] (c);
\vertex[above=3cm of b] (e);
\diagram*{
    (a)-- [scalar, edge label=\(t_2\hspace{10mm}t\)] (b),
    (b)-- [scalar, half left, edge label=\(\)] (c),
    (b)-- [plain, edge label=\(\hspace{10mm}t_1\)] (c),
    (b)-- [scalar, half right, edge label=\(\)] (c),
};
\end{feynman}
\end{tikzpicture}}_\text{Order 2, type C}
+\underbrace{\begin{tikzpicture}
\begin{feynman}[inline=(a)]
\vertex(b);
\vertex[right=2cm of a] (b);
\vertex[above= of b] (c);
\vertex[above= of c](d);
\diagram*{
    (a)-- [scalar, edge label=\(t_2\hspace{10mm}\)] (b),
    (b)-- [scalar, half left, edge label=\(t_1\)] (c),
    (c)-- [scalar, half left, edge label=\(t\)] (d),
    (d)-- [scalar, half left, edge label=\(\)] (c),
    (c)-- [plain, half left, edge label=\(\)] (b),
};
\end{feynman}
\end{tikzpicture}}_\text{Order 2, type A}
+\underbrace{\begin{tikzpicture}
\begin{feynman}[inline=(a)]
\vertex(b);
\vertex[right=2cm of a] (b);
\vertex[above= of b] (c);
\vertex[above= of c](d);
\vertex[right= of b] (e);
\diagram*{
    (a)-- [plain, edge label=\(t_2\hspace{10mm}\)] (b),
    (b)-- [plain, half left, edge label=\(t'\)] (c),
    (c)-- [scalar, half left, edge label=\(t\)] (d),
    (d)-- [scalar, half left, edge label=\(\)] (c),
    (c)-- [scalar, half left, edge label=\(\)] (b),
    (b)-- [plain, edge label=\(\hspace{10mm}t_1\)] (e),
};
\end{feynman}
\end{tikzpicture}}_\text{Order 2, type RA}
\end{eqnarray}
\begin{eqnarray}
\label{ExpDiagrams2}
\hat{\mathcal{L}}^{l_1}_{\mathbf{p}_1, t_1}\begin{tikzpicture}
\begin{feynman}[inline=(a)]
\vertex(b);
\vertex[right=2cm of b] (d);
\diagram*{
    (b)-- [photon, edge label=\(t_2\hspace{10mm}t_1\)] (d),
};
\end{feynman}
\end{tikzpicture}
&=&
\underbrace{\delta_{l_1,l_2}\delta(\mathbf{p}_1-\mathbf{p}_2)\delta(t_1-t_2)}_\text{Order 0}+
\underbrace{\begin{tikzpicture}
\begin{feynman}[inline=(e)]
\vertex(a);
\vertex[right=2cm of a] (b);
\vertex[above= of b] (c);
\diagram*{
    (a)-- [plain, edge label=\(t_2\hspace{15mm}\)] (b),
    (b)-- [scalar, half left, edge label=\(t_1\)] (c),
    (c)-- [scalar, half left, edge label=\(\)] (b),
};
\end{feynman}
\end{tikzpicture}}_\text{Order 1, type RA}\\
&&+\underbrace{\begin{tikzpicture}
\begin{feynman}[inline=(a.base)]
\vertex(a);
\vertex[right=2cm of a] (b);
\vertex[above= of b] (c);
\vertex[right=2cm of b] (d);
\vertex[above= of d] (e);
\diagram*{
    (a)-- [plain, edge label=\(t_0\hspace{15mm}\)] (b),
    (b)-- [scalar, half left, edge label=\(t\)] (c),
    (c)-- [scalar, half left, edge label=\(\)] (b),
    (b)-- [plain, edge label=\(\hspace{18mm}\)] (d),
    (d)-- [scalar, half left, edge label=\(t_1\)] (e),
    (e)-- [scalar, half left, edge label=\(\)] (d),
};
\end{feynman}
\end{tikzpicture}}_\text{Order 2, type RA}
+\underbrace{\begin{tikzpicture}
\begin{feynman}[inline=(a)]
\vertex(b);
\vertex[right=2cm of a] (b);
\vertex[above= of b] (c);
\vertex[above= of c](d);
\vertex[right= of b] (e);
\diagram*{
    (a)-- [plain, edge label=\(t_2\hspace{10mm}\)] (b),
    (b)-- [plain, half left, edge label=\(t_1\)] (c),
    (c)-- [scalar, half left, edge label=\(t\)] (d),
    (d)-- [scalar, half left, edge label=\(\)] (c),
    (c)-- [scalar, half left, edge label=\(\)] (b),
};
\end{feynman}
\end{tikzpicture}}_\text{Order 2, type RA}\nonumber\\
&&+
\underbrace{\begin{tikzpicture}[baseline=(a)]
\begin{feynman}[inline=(a)]
\vertex(a);
\vertex[right=2cm of a] (b);
\vertex[right=2cm of b] (c);
\vertex[above=3cm of b] (e);
\diagram*{
    (a)-- [plain, edge label=\(t_2\hspace{10mm}t\)] (b),
    (b)-- [scalar, half left, edge label=\(\)] (c),
    (b)-- [plain, edge label=\(\hspace{10mm}t_1\)] (c),
    (b)-- [scalar, half right, edge label=\(\)] (c),
};
\end{feynman}
\end{tikzpicture}}_\text{Order 2, type RB}\nonumber
\end{eqnarray}
\begin{multicols}{2}
Here, the single \textit{loops} means that there is a 0-order correlation function evaluated  at equal times. Also, the corresponding order of each diagram is associated with the number of vertices. Concurrently, this number indicates the number of  coefficient factors $L^{ss_1s_2s_3}_{\mathbf{k}\mathbf{k}_1\mathbf{k}_2\mathbf{k}_3}$ involved in wave vector integrals. Superscript sums and combinatory factors are ignored in the diagramatical representation.\\

To obtain the DIA equations, a resummation scheme is necessary. Such  resummation process consists of considering an infinite subset of diagrams, to capture some effects beyond the weakly nonlinear regime. In practice, the resummation corresponds to replace the free correlations and free response function  ( $C_0$ or $R_0$) on the right side of equation (\ref{ExpDiagrams}) and (\ref{ExpDiagrams2}) with the respective \textit{"exact"} correlation and response function, i.e, $C_0\rightarrow C$ and $R_0\rightarrow R$ for each mode. Finally, we obtain  the DIA equations, for the two point correlation
\begin{eqnarray}
\label{DIADiagramsC}
\hat{\mathcal{L}}^{l_1}_{p_1, t_1}
\begin{tikzpicture}
\begin{feynman}[inline=(a)]
\vertex(b);
\vertex[right=2cm of b] (d);
\diagram*{
    (b)-- [gluon, edge label=\(t_2\hspace{15mm}t_1\)] (d),
};
\end{feynman}
\end{tikzpicture}&=&
\underbrace{\begin{tikzpicture}
\begin{feynman}[inline=(a)]
\vertex(b);
\vertex[right=2cm of b] (d);
\diagram*{
    (b)-- [photon, edge label=\(t_2\hspace{15mm}t_1\)] (d),
};
\end{feynman}
\end{tikzpicture}}_\text{Type R}
+
\underbrace{\begin{tikzpicture}
\begin{feynman}[inline=(b)]
\vertex(a);
\vertex[right=2cm of a] (b);
\vertex[above= of b] (c);
\diagram*{
    (a)-- [gluon, edge label=\(t_2\hspace{15mm}\)] (b),
    (b)-- [gluon, half left, edge label=\(t_1\)] (c),
    (c)-- [gluon, half left, edge label=\(\)] (b),
};
\end{feynman}
\end{tikzpicture}}_\text{Type A}
\nonumber\\
&&+\underbrace{\begin{tikzpicture}[baseline=(a)]
\begin{feynman}[inline=(a)]
\vertex(a);
\vertex[right=2cm of a] (b);
\vertex[right=2cm of b] (c);
\vertex[above=3cm of b] (e);
\diagram*{
    (a)-- [photon, edge label=\(t_2\hspace{13mm}t\)] (b),
    (b)-- [gluon, half left, edge label=\(\)] (c),
    (b)-- [gluon, edge label=\(\hspace{13mm}t_1\)] (c),
    (b)-- [gluon, half right, edge label=\(\)] (c),
};
\end{feynman}
\end{tikzpicture}}_\text{Type B}
\nonumber
\\
&&+
\underbrace{\begin{tikzpicture}[baseline=(a)]
\begin{feynman}[inline=(a)]
\vertex(a);
\vertex[right=2cm of a] (b);
\vertex[right=2cm of b] (c);
\vertex[above=3cm of b] (e);
\diagram*{
    (a)-- [gluon, edge label=\(t_2\hspace{13mm}t\)] (b),
    (b)-- [gluon, half left, edge label=\(\)] (c),
    (b)-- [photon, edge label=\(\hspace{13mm}t_1\)] (c),
    (b)-- [gluon, half right, edge label=\(\)] (c),
};
\end{feynman}
\end{tikzpicture}}_\text{Type C}
\nonumber\\
\end{eqnarray}
and for the response function
\begin{eqnarray}
\hat{\mathcal{L}}^{l_1}_{p_1, t_1}
\begin{tikzpicture}
\begin{feynman}[inline=(a)]
\vertex(b);
\vertex[right=2cm of b] (d);
\diagram*{
    (b)-- [photon, edge label=\(t_2\hspace{15mm}t_1\)] (d),
};
\end{feynman}
\end{tikzpicture}&=&
\delta_{l_1,l_2}\delta(t_1-t_2)+
\underbrace{\begin{tikzpicture}
\begin{feynman}[inline=(e)]
\vertex(a);
\vertex[right=2cm of a] (b);
\vertex[above= of b] (c);
\diagram*{
    (a)-- [photon, edge label=\(t_2\hspace{15mm}\)] (b),
    (b)-- [gluon, half left, edge label=\(t_1\)] (c),
    (c)-- [gluon, half left, edge label=\(\)] (b),
};
\end{feynman}
\end{tikzpicture}}_\text{Type RA}
\nonumber\\
&&+
\underbrace{\begin{tikzpicture}[baseline=(a)]
\begin{feynman}[inline=(a)]
\vertex(a);
\vertex[right=2cm of a] (b);
\vertex[right=2cm of b] (c);
\vertex[above=3cm of b] (e);
\diagram*{
    (a)-- [photon, edge label=\(t_2\hspace{13mm}t'\)] (b),
    (b)-- [gluon, half left, edge label=\(\)] (c),
    (b)-- [photon, edge label=\(\hspace{13mm}t_1\)] (c),
    (b)-- [gluon, half right, edge label=\(\)] (c),
};
\end{feynman}
\end{tikzpicture}}_\text{Type RB}
\nonumber\\
\end{eqnarray}
It is important to note that the diagrams in the final DIA equations corresponds to an infinite subset of the original expansion (\ref{ExpDiagrams}). One can show that all the diagrams up to the second order in (\ref{ExpDiagrams})  are considered in the DIA equations, but also many more (see for details  Appendix \ref{B2} ).  For instance, one can easily notice that diagrams of order 1 and 2 of  type A in (\ref{ExpDiagrams}) can be recovered from the type A diagram in the DIA equation (\ref{DIADiagramsC}). However, diagrams of order 2 type B and C in (\ref{ExpDiagrams}) are not obtained  from the type A diagram in the DIA equation (\ref{DIADiagramsC}), but from diagrams type B and C in (\ref{DIADiagramsC}) (see  Appendix \ref{B2} for details).\\

Considering statistical spatial homogeneity, the correlation and response function can be simplified as   $C^{l_1l_2}_{\mathbf{p_1}\mathbf{p}_2}(t_1,t_2)= \delta(\mathbf{p}_1+\mathbf{p}_2) C^{l_1l_2}_{\mathbf{p}_2}(t_1,t_2)2\pi$ and $R^{l_1l_2}_{\mathbf{p}_1\mathbf{p}_2}(t_1,t_2)=\delta(\mathbf{p}_1+\mathbf{p}_2)R^{l_1l_2}_{\mathbf{p}_2}(t_1,t_2)$. Then the DIA equations read as
\begin{small}
\begin{eqnarray}
\label{DIAC}
\hat{\mathcal{L}}^{l_1}_{\mathbf{p_2},t_1}C^{l_1l_2}_{\mathbf{p}_2}(t_1,t_2)&=&\sum_{s}f^{l_1s}_{\bb{p}_2}C^{sl_2}_{\mathbf{p}_2}(t_1,t_2)+2F_{\mathbf{p}_2}R^{l_2l_1}_{\mathbf{p}_2}(t_2,t_1)\nonumber\\
&&+\int^{t_1}_{0}d\tau\sum_{x_3}\sigma^{l_1x_3}_{\bb{p}_2}(t_1,\tau)C^{x_3l_2}_{\mathbf{p}_2}(\tau,t_2)\nonumber\\
&&+\int^{t_2}_{0}d\tau \sum_{x}\mathcal{S}^{l_1x}_{\bb{p}_2}(\tau,t_1)R^{l_2-x}_{\bb{p}_2}(t_2,\tau)\nonumber\\
\end{eqnarray}
\end{small}
\begin{small}
\begin{eqnarray}
\label{DIAR}
\hat{\mathcal{L}}^{l_1}_{\mathbf{p_2},t_1}R^{l_1-l_2}_{-\mathbf{p}_2}(t_1,t_2)&=&\delta_{l_1,l_2}\delta(t_1-t_2)+\sum_{s}f^{l_1s}_{\bb{p}_2}R^{s-l_2}_{-\mathbf{p}_2}(t_1,t_2)\nonumber\\
&&+\int^{t_1}_{t_2}d\tau\sum_{x_3}\sigma^{l_1x_3}_{-\bb{p}_2}(t_1,\tau)R^{x_3-l_2}_{-\mathbf{p}_2}(\tau,t_2)\nonumber\\
\end{eqnarray}
\end{small}
with
\begin{small}
\begin{eqnarray}
f^{ls}_{\bb{p}}(t)&:=&3(2\pi)\sum_{s_1s_2}\int d\bb{k} L^{ls_1s_2s}_{\bb{-pkk-p}}C^{s_1s_2}_{\bb{k}}(t,t)\\
\sigma^{l_1x_3}_{\bb{p}}(t_1,t)&:=& 18(2\pi)^2\sum_{s_1s_2s_3} \int d\mathbf{k}_{123}L^{l_1s_1s_2s_3}_{-\mathbf{p}_2\mathbf{k}_1\mathbf{k}_2\mathbf{k}_3} \nonumber\\
&&\times\sum_{xx_1x_2} L^{xx_1x_2x_3}_{\mathbf{k}_1-\mathbf{k}_3-\mathbf{k}_2-\mathbf{p}_2}C^{x_1s_3}_{\mathbf{k}_3}(t,t_1)C^{x_2s_2}_{\mathbf{k}_2}(t,t_1)\nonumber\\
&&\times  R^{s_1-x}_{\mathbf{k}_1}(t_1,t)\delta(\mathbf{k}_1+\mathbf{k}_2+\mathbf{k}_3+\mathbf{p}_2)\\
\mathcal{S}^{l_1x}_{\bb{p}_2}(t,t_1)&:=&6(2\pi)^2\sum_{s_1s_2s_3} \int d\mathbf{k}_{123}L^{l_1s_1s_2s_3}_{-\mathbf{p}_2\mathbf{k}_1\mathbf{k}_2\mathbf{k}_3} \nonumber\\
&&\times\sum_{x_1x_2x_3} L^{xx_1x_2x_3}_{\mathbf{p}_2-\mathbf{k}_1-\mathbf{k}_2-\mathbf{k}_3}C^{x_1s_1}_{\mathbf{k}_1}(t,t_1)C^{x_2s_2}_{\mathbf{k}_2}(t,t_1)\nonumber\\
&&\times C^{x_3s_3}_{\mathbf{k}_3}(t,t_1)  \delta(\mathbf{k}_1+\mathbf{k}_2+\mathbf{k}_3+\mathbf{p}_2).
\end{eqnarray}
\end{small} 

Equations (\ref{DIAC}) and (\ref{DIAR}) correspond to the DIA equations for the elastic plate, similar to the DIA equations originally derived by Kraichnan in \cite{kraichnan1959structure} for hydrodynamic turbulence. The main difference lies in the one-loop contribution proportional to $f^{l_1s}_{\bb{p}_2}$, observed on the right hand side of equation (\ref{DIAC}) and (\ref{DIAR}). As one could notice, this nonlinear term alone is not capable for inducing an energy flux toward the small scales. If the amplitudes were zero above some wave number at some initial time, only the two-loop terms can activate those modes.  However, the one-loop term could be very relevant for phase mixing and the decorrelation time scale in the system, which can be responsible for the break-down of the weakly nonlinear regime.  As one would anticipate, this term will be shown to be responsible for the frequency shift in the weakly nonlinear regime.

\section{DIA Properties}
\subsection{Underlying stochastic system}
An important property of the DIA equations is that they correspond to an exact closure for an underlying stochastic system. This ensures the existence of a well-defined PDF and quantities like $C^{l_1-l_1}_{\bf{k}}(t,t)$ are positive definite. Constructing backwards from equation (\ref{DIAC}) we get the stochastic system\footnote{The derivation from the stochastic system to the DIA equations is shown in Appendix \ref{E}}:
\begin{small}
\begin{eqnarray}
\label{Stocsys}
(\partial_t+il\omega_{\bb{p}})A^{l}_{\bb{p}}(t)-\sum_{s}f^{ls}_{\bb{p}}A^{s}_{\bb{p}}(t)-\int^{t}_{0}\sum_{x}\sigma^{lx}_{\bb{p}}(t_1,\tau)A^{x}_{\bb{p}}(\tau)d\tau=\eta^{l}_{\bb{p}}(t)\nonumber\\
\end{eqnarray}
\end{small}
where $\eta^{l}_{\bb{p}}(t)$ is the stochastic force which is statistical independent among modes and must satisfy $\langle \eta^{l}_{\bb{p}}(t)\eta^{x}_{\bb{p}_2}(t')\rangle=\mathcal{S}^{lx}_{\bb{p}}(t,t')2\pi\delta(\bb{p+p_2})$. A simple way to accomplish this constraint is defining
\begin{small}
\begin{eqnarray}
\label{noise2}
\eta^{l}_{\bb{p}}(t)&:=&\sqrt{6(2\pi)^3}\sum_{s_1s_2s_3}\int d\bb{k}_{123}L^{ls_1s_2s_3}_{\bb{-pk_1k_2k_3}}\nonumber\\
&&\times\xi^{s_1}_{\bb{k}_1}(t)\chi^{s_2}_{\bb{k}_2}(t)\phi^{s_3}_{\bb{k}_3}(t)\delta(\bb{k_1+k_2+k_3+p})
\end{eqnarray}
\end{small}
where $\xi^{s_1}_{\bb{k}_1}(t),\chi^{s_2}_{\bb{k}_2}(t)$ and $\phi^{s_3}_{\bb{k}_3}(t)$ are independent complex stochastic variables which correlation $\langle\xi^{s_1}_{\bb{k}_1}(t_1)\xi^{s_2}_{\bb{k}_2}(t_2)\rangle=\langle\chi^{s_1}_{\bb{k}_1}(t_1)\chi^{s_2}_{\bb{k}_2}(t_2)\rangle=\langle\phi^{s_1}_{\bb{k}_1}(t_1)\phi^{s_2}_{\bb{k}_2}(t_2)\rangle=\delta(\bb{k_1+k_2})C^{s_1s_2}_{\bb{k}_2}(t_1,t_2)$.
Therefore, the correlation and response functions of the SDE (\ref{Stocsys}) under the noise (\ref{noise2}) are precisely the ones given by the DIA equations.

\subsection{Kinetic equation}
It is important to note that the DIA contains all the diagrams related to the second order expansion and therefore, contains all the needed diagrams to derive the kinetic equation of wave turbulence. If we want to take the weakly nonlinear limit, we have to rescale the order of the deformation, $A^s_{\mathbf{k}}\rightarrow \epsilon A^s_{\mathbf{k}}$, the forcing $F_{\mathbf{p}}\rightarrow\epsilon^2F_{\mathbf{p}}$ and dissipation $\gamma_{\mathbf{p}}\rightarrow\epsilon^2\gamma_{\mathbf{p}}$. It is convenient to make a change of variables to the  time difference : $(t_1,t_2)\rightarrow(\tau,t_2)$ with $\tau\coloneqq t_2-t_1$.  A standard perturbation method leads at zero order to 
\begin{eqnarray}
 C^{l_1l_2}_{0,\mathbf{p}_2}(\tau,t_2)&=&C^{l_1l_2}_{0,\mathbf{p}_2}(0,0)e^{-i(l_1+l_2)\omega_{\mathbf{p}_2}t_2+il_1\omega_{\mathbf{p}_2}\tau}\nonumber\\
 R^{l_1l_2}_{0,\mathbf{p}_2}(\tau)&=&\delta_{l_1,-l_2}e^{-il_1\omega_{\bb{p_2}}\tau}\nonumber
 \end{eqnarray}
which are the same as Eq.\ref{C0} and  Eq.\ref{R0}  but without injection and dissipation. At higher orders of the expansion, resonant terms emerge requiring a multiscale perturbation method. Keeping the time difference $\tau$ as a fast time scale of order one and considering the long time behavior of $t_2$, one can rewrite  $C^{l_1l_2}_{\mathbf{p}_2}(\tau,t_2)=C^{l_1l_2}_{\mathbf{p}_2,0}(\tau,t_2,T_2,\mathcal{T}_2)+\epsilon C^{l_1l_2}_{\mathbf{p}_2,1}(\tau,t_2,T_2,\mathcal{T}_2)+\epsilon^2 C^{l_1l_2}_{\mathbf{p}_2,2}(\tau,t_2,T_2,\mathcal{T}_2)+\ldots$ with $T_2:=\epsilon t_2$ and $\mathcal{T}_2\coloneqq\epsilon^2t_2$. The integration constant of the zero order perturbation can then depend on the slow time, thus $C^{l_1l_2}_{0,\mathbf{p}_2}(0,0)\rightarrow C^{l_1l_2}_{0,\mathbf{p}_2}(0,0,T_2,\mathcal{T}_2)$. The slow time evolution  of the correlation function $C^{l_1l_2}_{0,\mathbf{p}_2}(0,0,T_2,\mathcal{T}_2)$ is then set by the secular condition necessary to ensure an asymptotic  perturbation expansion. At the second order expansion, one obtains that the two point correlation function at $\tau=0$ defined as $n_{l_2\mathbf{p}_2}(t_2):=C^{l_2-l_2}_{0,\mathbf{p}_2}(0,0,0,\mathcal{T}_2)$ must satisfy the well known kinetic equation for elastic plates \cite{during2006weak}(see Appendix  \ref{C} for details):
\begin{small}
\begin{eqnarray}
\frac{dn_{l_2\bb{p_2}}}{dt}&=&12l_2\pi\epsilon^4\int d\bb{k}_{123} |L^{l_2s_1s_2s_3}_{\bb{p_2k_1k_2k_3}}|^2\sum_{s_1s_2s_3}n_{s_1\bb{k}_1}n_{s_2\bb{k}_2}n_{s_3\bb{k}_3}n_{l_2\bb{p_2}}\nonumber\\
&&\times\left(\frac{l_2}{n_{l_2\bb{p_2}}}-\frac{s_1}{n_{s_1\bb{k}_1}}-\frac{s_2}{n_{s_2\bb{k}_2}}-\frac{s_3}{n_{s_3\bb{k}_3}}\right)\nonumber\\
&&\times \delta(l_2\omega_{p_2}-s_1\omega_{k_1}-s_2\omega_{k_2}-s_3\omega_{k_3})\nonumber\\
&&\times\delta(\bb{k}_1+\bb{k}_2+\bb{k}_3-\bb{p_2})\nonumber\\
\end{eqnarray}
\end{small}
It should be noted that although the derivation of DIA equations does not require continuum variables in Fourier space, the derivation of the kinetic equation does (see Appendix  \ref{C}).  

\section{Discussion}

The Martin-Siggia-Rose path integral formalism used in this work gives a natural and clear approach to derive DIA and study wave turbulence theory for the elastic plates and other systems. This simplicity arises from the explicit dependence of the path integral on the response function, which enables a straightforward derivation of DIA. Interestingly, a new term emerges in the DIA equations for elastic plates, which is not present in previously studied systems. It originated from one-loop corrections,  but yet, its physical significance and its role in non-weak turbulent systems remain to be explored. 

 Since the weak turbulence theory is recovered from the DIA equations in the proper limit, we expect the present formalism will elucidate several open questions on Wave Turbulence. For instance, a description of the response function at the slow time scales of the kinetic equation or a better understanding of the full probability distribution function using the stochastic underlying equations in the weakly nonlinear regime. Since DIA equations are equally derived for finite size systems, questions regarding discreteness in wave turbulence could also be addressed. Finally and more interestingly, the DIA equations might shed light on the breakdown of wave turbulence and the critical balance regime \cite{newell2008role}. 

Theoretically, the DIA equations enable us to compute the so-called transport power \cite{kraichnan1959structure,kraichnan1964decay}. This is the rate at which energy is transferred from modes $k'$ below some particular $k$ ($k'<k$) to modes above the same $k$ ($k'>k$).Therefore, we propose to study the evolution of this quantity for different amplitudes and consequently, attempt to categorize regimes according to the transport power.\\

Finally, it should be noted that for incompressible fluids using the Clebsch variable, similar DIA equations are obtained. Since no small parameter exists for such a system, the validity of DIA  is questionable but it could be interesting to establish wether it leads to the same incorrect results obtained by Kraichnan for the DIA in fluids  \cite{kraichnan1959structure}.

\textbf{ Acknowledgment.} We thank G.Krstulovic, S. Rica and C.Falcón, for fruitful discussions and for comments on the manuscript.  I.P and G.D. acknowledge support from Fondecyt Grant No. $1210656$ and FONDECYT Grant No.$1181382$.

\bibliographystyle{apsrev4-1}

\bibliography{PaperBib}

\cleardoublepage

\appendix
\section{Generating Functional}
\label{A}
 To obtain the solution for the generating functional at order 0, we start with the linear version of the path integral. As we would like to have a quadratic form in the argument of the exponential, we write $S_0$ ( the linear version of MSR action $S$) as:
 \begin{equation}
 S_0= \frac{1}{2}\sum_{s_1s_2}\int d\bb{k_1}\int d\bb{k}_2\int\int \phi^{s_1,T}_{\bb{k}_1}(t)G^{-1,s_1s_2}_{\bb{k_1k_2}}(t,t')\phi^{s_2}_{\bb{k_2}}(t')dtdt'
 \end{equation}
 where $G^{-1,s_1s_2}_{\bb{k_1k_2}}(t,t'):=\begin{pmatrix} G^{-1,s_1s_2}_{\bb{k_1k_2},1}(t,t') & G^{-1,s_1s_2}_{\bb{k_1k_2},2}(t,t')\\G^{-1s_1s_2}_{\bb{k_1k_2},3}(t,t') & G^{-1s_1s_2}_{\bb{k_1k_2},4}(t,t') \end{pmatrix}$, $\phi^{s}_{\bb{k}}(t):= \begin{pmatrix} A^{s}_{\bb{k}}(t)\\ \tilde{A}^{s}_{\bb{k}} (t)\end{pmatrix}$ and $\phi^{s,T}_{\bb{k}}(t)$ is the transpose of $\phi^{s}_{\bb{k}}(t)$.  Then, by comparison we find out that:
 \begin{eqnarray}
 G^{-1,s_1s_2}_{\bb{k_1k_2},1}(t,t')&=&0\nonumber\\
 G^{-1,s_1s_2}_{\bb{k_1k_2},2}(t,t')&=&\delta(t-t')\delta_{s_1,s_2}\delta(\bb{k_1}-\bb{k_2})(-\partial_{t'}+is_2\Omega^{s_2}_{-\bb{k_2}})\nonumber\\
 G^{-1s_1s_2}_{\bb{k_1k_2},3}(t,t') &=& \delta(t-t')\delta_{s_1,s_2}\delta(\bb{k_1}-\bb{k_2})(\partial_{t'}+is_2\Omega^{s_2}_{\bb{k_2}})\nonumber\\
  G^{-1s_1s_2}_{\bb{k_1k_2},4}(t,t')&=& -2F_{\bb{k_2}}\delta(\bb{k_1}+\bb{k_2})\delta_{s_1,-s_2}\delta(t-t')
  \end{eqnarray}
 As already mentioned, here we add the auxiliar fields compressed in $j^s_{\bb{k}}(t)$. So we get:
 \begin{small}
 \begin{eqnarray}
 \label{Z00}
Z_0[J,\tilde{J}]&=&\int \mathcal{D}[A^{s}_{\bb{k}}(t)]\mathcal{D}[\tilde{A}^{s}_{\bb{k}}(t)]\nonumber\\
&&\times e^{-\frac{1}{2}\sum_{s_1s_2}\int d\bb{k}_1\int d\bb{k}_2\int^T_0\int^T_0 dtdt' \phi^{s_1,T}_{\bb{k}_1}(t)G^{-1,s_1s_2}_{\bb{k_1k_2}}(t,t')\phi^{s_2}_{\bb{k}_2}(t')}\nonumber\\
&&\times e^{\sum_{s_1}\int d\bb{k}_1\int j^{s_1,T}_{\bb{k}_1}(t)\phi^{s_1}_{\bb{k}_1}(t)dt}\nonumber\\
\end{eqnarray}
\end{small} 
As usual, we now follow the common procedure to integrate quadratic forms. That is, we first make a translation of variable $y^s_{\bb{k}}(t):=\phi^s_{\bb{k}}(t)-\phi^{*s}_{\bb{k}}(t)$ where $\left[\frac{\delta S_0}{\delta \phi^{s}_{\bb{k}}(t)}\right]_{\phi=\phi^{*}}=0$. From this last condition we get: 
\begin{equation}
\label{phiestrella}
\sum_s\int d\bb{k}\int^T_0 dt' G^{-1,s_1s}_{\bb{k_1k}}(t,t')\phi^{*s}_{\bb{k}}(t')=j^{s_1}_{\bb{k}_1}(t)
\end{equation}
Thus, to obtain $\phi^{*s}_{\bb{k}}(t')$, we need to invert this equation and that means we need to have the inverse of $G^{-1,s_1s}_{\bb{k_1k}}(t,t')$:
\begin{small}
\begin{equation}
\sum_{s_3}\int d\bb{k}_3\int d\tau G^{-1,s_1s_3}_{\bb{k_1k_3}}(t,\tau)G^{s_3s_2}_{\bb{k_3k_2}}(\tau,t')=\delta_{s_1s_2}\delta(\bb{k}_1-\bb{k}_2)\delta(t-t')\mathbb{I}_{2x2}
\end{equation}
\end{small}
This condition yields to equations for the components and we know functions that satisfy them: The correlation function and the response function, both at order 0. From the linear version of the dynamic equation for the fields, we know that:
\begin{equation}
(\partial_t+is_1\Omega^{s_1}_{\bb{k}_1})C^{s_1s_2}_{0,\bb{k_1k_2}}(t,t')=2F_{\bb{k}_1}R^{s_2s_1}_{0,\bb{k_2k_1}}(t',t)
\end{equation}
and
\begin{equation}
(\partial_t+is_1\Omega^{s_1}_{\bb{k}_1})R^{s_1-s_2}_{0,\bb{k_1-k_2}}(t,t')=\delta_{s_1s_2}\delta(\bb{k}_1-\bb{k}_2)\delta(t-t')
\end{equation}
Then we conclude that a posible solution is:
\begin{equation}
 \label{G}
G^{s_1s_2}_{\bb{k_1k_2}}(t,t')=
\begin{pmatrix}
C^{s_1s_2}_{0,\bb{k_1k_2}}(t,t') & R^{s_1-s_2}_{0,\bb{k_1-k_2}}(t,t')\\
R^{s_2-s_1}_{0,\bb{k_2-k_1}}(t',t) & 0
\end{pmatrix}
\end{equation}
Therefore, we can invert (\ref{phiestrella}) and make the change of variable. Finally, to diagonalize the matrix, we make a unitary rotation which yields a gaussian integral. Because of the unitary rotation and the linear translation, the jacobian for the variable change is 1. Finally, after integration, we get:
 \begin{eqnarray}
Z_0[J,\tilde{J}]= e^{-\frac{1}{2}\sum_{s_1}\int d\mathbf{k_1}\int dt j^{s_1,T}_{\mathbf{k_1}}(t)\phi^{*s_1}_{\mathbf{k_1}}(t)}
\end{eqnarray}
\section{Obtaining DIA}
\label{B}
Since we already know the generating functional for the linear case, we start from the definition of the exact generating functional and separate $S=S_0+S_{N.L}$:
\begin{small}
\begin{eqnarray}
Z_0&=& \int \mathcal{D}A^s_{\mathbf{k}}(t) \mathcal{D}\tilde{A}^s_{\mathbf{k}}(t)e^{-S_0[A^{s}_{\mathbf{k}}(t),\tilde{A}^{s}_{\mathbf{k}}(t)]-S_{N.L}[A^{s}_{\mathbf{k}}(t),\tilde{A}^{s}_{\mathbf{k}}(t)]}
 \end{eqnarray}
\end{small}
We then convert it into a power series and add the auxiliar fields:
\begin{small}
\begin{eqnarray}
Z[J,\tilde{J}]&=&\int \mathcal{D}[A^{s}_{\mathbf{k}}(t)\mathcal{D}[\tilde{A}^{s}_{\mathbf{k}}(t)]\sum^{\infty}_{n=0}\frac{1}{n!}\left(S_{N.L}\right)^n\nonumber\\
&&\times \exp\left( -S_0+\sum_s\int d\mathbf{k}\int \tilde{J}^s_{\mathbf{k}}(t)A^{s}_{\mathbf{k}}(t)+J^s_{\mathbf{k}}(t)\tilde{A}^{s}_{\mathbf{k}}(t)dt\right)\nonumber\\
\end{eqnarray}
\end{small} 
  If we expand $(S_{N.L})^n$, we observe that each term can be obtained if we perform derivatives with respect to the auxiliar fields and then set them to 0:
\begin{small}
\begin{eqnarray}
&&Z[J,\tilde{J}]\\
&=&\int \mathcal{D}[A^{s}_{k}(t)]\mathcal{D}[\tilde{A}^{s}_{k}(t)]\Bigg(1+\int d\mathbf{k}\int dt \sum_{ss_1s_2s_3}\int L^{ss_1s_2s_3}_{\mathbf{kk_1k_2k_3}}\nonumber\\
&&\times \tilde{A}^{s}_{\mathbf{k}}(t)A^{s_1}_{\mathbf{k_1}}(t)A^{s_2}_{\mathbf{k_2}}(t)A^{s_3}_{\mathbf{k_3}}(t)\delta (\mathbf{K}_{\mathbf{-k}})d\mathbf{k}_{123}\nonumber+O(2)\Bigg)\nonumber\\
&&\times \exp\left( -S_0+\sum_s\int d\mathbf{k} \int \tilde{J}^s_{\mathbf{k}}(t)A^{s}_{\mathbf{k}}(t)+J^s_{\mathbf{k}}(t)\tilde{A}^{s}_{\mathbf{k}}(t)dt\right)\nonumber\\
&=& \int \mathcal{D}[A^{s}_{\mathbf{k}}(t)]\mathcal{D}[\tilde{A}^{s}_{\mathbf{k}}(t)]\Bigg(1+\int d\mathbf{k}\int dt \sum_{ss_1s_2s_3}\int L^{ss_1s_2s_3}_{\mathbf{kk_1k_2k_3}}\nonumber\\
&&\times\frac{\delta^4}{\delta \tilde{J}^{s_3}_{\mathbf{k}_3}(t)\tilde{J}^{s_2}_{\mathbf{k}_2}(t)\tilde{J}^{s_1}_{\mathbf{k}_1}(t)J^{s}_{\mathbf{k}}(t)}\delta (\mathbf{K}_{\mathbf{-k}})d\mathbf{k}_{123}+O(2)\Bigg)\nonumber\\
&&\times \exp\left( -S_0+\sum_s\int d\mathbf{k} \int \tilde{J}^s_{\mathbf{k}}(t)A^{s}_{\mathbf{k}}(t)+J^s_{\mathbf{k}}(t)\tilde{A}^{s}_{\mathbf{k}}(t)dt\right)\nonumber
 \end{eqnarray}
\end{small}
where $\delta(\mathbf{K}_{\mathbf{-k}}):=\delta(\mathbf{k_1+k_2+k_3-k})$. Since the integrals are over the $A$ and $\tilde{A}$ fields, the series pass out the integrals and $Z$ can be viewed as a series of terms, each of them involving derivatives of free case $Z_0$ with respect to the auxiliar fields. Finally, to obtain the DIA, we go up to second order of the expansion and compute 10 derivatives (8 for the nonlinear expression of $Z_2$ and 2 to obtain either the correlation function or the response function). Nevertheless, instead of actually computing the 10 derivatives, we can make use of the diagram technique as follows. We need second order terms, which means that we will have 2 internal vertices corresponding to the 2 factors $L^{ss_1s_2s_3}_{\mathbf{kk_1k_2k_3}}$ and 2 external legs associated with the 2 times where the correlation/response function is actually evaluated. In this case, one of the external legs will have the time $t_2$ and the other $t_1$. Also, the 2 internal vertices correspond to 2 time variables (say $t$ and $t'$) which, obviously, are integrated. Next, we observe the symmetries that arise from the derivatives with respect to the $x,\mathbf{q}$ and $s,\mathbf{k}$, because they are all dummy variables. Since $Z_0$ is expressed in terms of the 0-order statistical functions and these functions come from differentiate twice with respect to the auxiliar fields as seen in \ ( for the $C^{l_1l_2}_{\mathbf{p_1p_2}}(t_1,t_2)$ the derivatives are $\frac{\delta^2}{\delta \tilde{J}^{l_2}_{\mathbf{p}_2}(t_2)\delta\tilde{J}^{l_1}_{\mathbf{p_1}}(t_1)}$, while for $R^{l_1l_2}_{\mathbf{p_1p_2}}(t_1,t_2)$ the derivatives are $\frac{\delta^2}{\delta J^{l_2}_{\mathbf{p}_2}(t_2)\delta\tilde{J}^{l_1}_{\mathbf{p_1}}(t_1)}$), we can view the problem of derivatives as a combinatory problem where we have to count every posible outcome that can be made if we take the derivatives by pairs. In this way, the only posible outcomes for the correlation function come from:
\begin{small}
 \begin{eqnarray}
 a)&&\left(\frac{\delta}{\delta \tilde{J}^{l_1}_{p_1}(t_1)}\frac{\delta}{\delta J^{s}_{k}(t)}\right) \left(\frac{\delta}{\delta \tilde{J}^{x_3}_{q_3}(t')}\frac{\delta}{\delta \tilde{J}^{x_2}_{q_2}(t')}\right) \left(\frac{\delta}{\delta \tilde{J}^{s_3}_{k_3}(t)}\frac{\delta}{\delta \tilde{J}^{s_2}_{k_2}(t)}\right) \nonumber\\
 &&\times\left(\frac{\delta}{\delta J^{x}_{q}(t')}\frac{\delta}{\delta \tilde{J}^{s_1}_{k_1}(t)}\right) \left(\frac{\delta}{\delta \tilde{J}^{l_2}_{p_2}(t_2)}\frac{\delta}{\delta \tilde{J}^{x_1}_{q_1}(t')}\right)
 \end{eqnarray}
 \end{small}
 which will represent:
 \begin{tikzpicture}
\begin{feynman}[inline=(a)]
\vertex(a);
\vertex[right=2cm of a] (b);
\vertex[above= of b] (c);
\vertex[right=2cm of b](d);
\vertex[above= of d] (e);
\vertex[right=2cm of d](f);
\diagram*{
    (a)-- [scalar, edge label=\(t_2\hspace{10mm}\)] (b),
    (b)-- [scalar, half left, edge label=\(t'\)] (c),
    (c)-- [scalar, half left, edge label=\(\)] (b),
    (b)-- [plain, edge label=\(\hspace{15mm}\)] (d),
    (d)-- [scalar, half left, edge label=\(t\)] (e),
    (e)-- [scalar, half left, edge label=\(\)] (d),
    (d)-- [plain, edge label=\(\hspace{15mm}t_1\)] (f),
};
\end{feynman}
\end{tikzpicture}
  and because of symmetry, there will be 18 diagrams of the same type.\\
\begin{small}
\begin{eqnarray}
b)&&\left(\frac{\delta}{\delta \tilde{J}^{l_1}_{p_1}(t_1)}\frac{\delta}{\delta J^{s}_{k}(t)}\right)\left(\frac{\delta}{\delta \tilde{J}^{x_3}_{q_3}(t')}\frac{\delta}{\delta \tilde{J}^{s_3}_{k_3}(t)}\right)\left(\frac{\delta}{\delta \tilde{J}^{x_2}_{q_2}(t')}\frac{\delta}{\delta \tilde{J}^{s_2}_{k_2}(t)}\right)\nonumber\\
&&\times\left(\frac{\delta}{\delta \tilde{J}^{l_2}_{p_2}(t_2)}\frac{\delta}{\delta \tilde{J}^{x_1}_{q_1}(t')}\right)\left(\frac{\delta}{\delta J^{x}_{q}(t')}\frac{\delta}{\delta \tilde{J}^{s_1}_{k_1}(t)}\right)
 \end{eqnarray}
 \end{small}
 represented by:\begin{tikzpicture}[baseline=(a)]
\begin{feynman}[inline=(a)]
\vertex(a);
\vertex[right=2cm of a] (b);
\vertex[right=2cm of b] (c);
\vertex[above=3cm of b] (e);
\vertex[right=2cm of c] (d);
\diagram*{
    (a)-- [scalar, edge label=\(t_2\hspace{10mm}t'\)] (b),
    (b)-- [scalar, half left, edge label=\(\)] (c),
    (b)-- [plain, edge label=\(\hspace{10mm}t\)] (c),
    (b)-- [scalar, half right, edge label=\(\)] (c),
    (c)-- [plain, edge label=\(\hspace{10mm}t_1\)] (d),
};
\end{feynman}
\end{tikzpicture}
 and there will be 36 of them.\\
 \begin{small}
\begin{eqnarray}
c)&&\left(\frac{\delta}{\delta \tilde{J}^{l_1}_{p_1}(t_1)}\frac{\delta}{\delta J^{s}_{k}(t)}\right)\left(\frac{\delta}{\delta \tilde{J}^{x_3}_{q_3}(t')}\frac{\delta}{\delta \tilde{J}^{s_3}_{k_3}(t)}\right)\left(\frac{\delta}{\delta \tilde{J}^{x_2}_{q_2}(t')}\frac{\delta}{\delta \tilde{J}^{s_2}_{k_2}(t)}\right)\nonumber\\
&&\times\left(\frac{\delta}{\delta \tilde{J}^{l_2}_{p_2}(t_2)}\frac{\delta}{\delta J^{x}_{q}(t') }\right)\left(\frac{\delta}{\delta \tilde{J}^{x_1}_{q_1}(t')}\frac{\delta}{\delta \tilde{J}^{s_1}_{k_1}(t)}\right)
 \end{eqnarray}
 \end{small}
 represented by:\begin{tikzpicture}[baseline=(a)]
\begin{feynman}[inline=(a)]
\vertex(a);
\vertex[right=2cm of a] (b);
\vertex[right=2cm of b] (c);
\vertex[above=3cm of b] (e);
\vertex[right=2cm of c] (d);
\diagram*{
    (a)-- [plain, edge label=\(t_2\hspace{10mm}t'\)] (b),
    (b)-- [scalar, half left, edge label=\(\)] (c),
    (b)-- [scalar, edge label=\(\hspace{10mm}t\)] (c),
    (b)-- [scalar, half right, edge label=\(\)] (c),
    (c)-- [plain, edge label=\(\hspace{10mm}t_1\)] (d),
};
\end{feynman}
\end{tikzpicture}
and there will be 12 of them.
\begin{small}
\begin{eqnarray}
d)&&\left(\frac{\delta}{\delta \tilde{J}^{l_1}_{p_1}(t_1)}\frac{\delta}{\delta J^{s}_{k}(t)}\right)\left(\frac{\delta}{\delta \tilde{J}^{x_3}_{q_3}(t')}\frac{\delta}{\delta \tilde{J}^{x_2}_{q_2}(t')}\right)\left(\frac{\delta}{\delta \tilde{J}^{x_1}_{q_1}(t')}\frac{\delta}{\delta \tilde{J}^{s_3}_{k_3}(t)}\right)\nonumber\\
&&\times \left(\frac{\delta}{\delta \tilde{J}^{l_2}_{p_2}(t_2)}\frac{\delta}{\delta \tilde{J}^{s_2}_{k_2}(t) }\right)\left(\frac{\delta}{\delta \tilde{J}^{x}_{q}(t')}\frac{\delta}{\delta \tilde{J}^{s_2}_{k_2}(t)}\right)
\end{eqnarray}
\end{small}
represented by:
\begin{tikzpicture}
\begin{feynman}[inline=(a)]
\vertex(b);
\vertex[right=2cm of a] (b);
\vertex[above= of b] (c);
\vertex[above= of c](d);
\vertex[right= of b] (e);
\diagram*{
    (a)-- [scalar, edge label=\(t_2\hspace{10mm}\)] (b),
    (b)-- [scalar, half left, edge label=\(t'\)] (c),
    (c)-- [scalar, half left, edge label=\(t\)] (d),
    (d)-- [scalar, half left, edge label=\(\)] (c),
    (c)-- [plain, half left, edge label=\(\)] (b),
    (b)-- [plain, edge label=\(\hspace{10mm}t_1\)] (e),
};
\end{feynman}
\end{tikzpicture}
with 18 of them. The procedure for the response function is completely analogue.\\
 It is important to clarify that since we are not assuming relations between $t_1$ and $t_2$ ( for now) , in the perturbation expansion , next to the non-symmetric-in-time diagrams (that is a,b and d) are the same diagrams but with $t_1$ and $t_2$ interchanged. This last argument will not hold for the response function, because of causality. Next, we apply the differential operator $\hat{\mathcal{L}}^{l_1}_{\mathbf{p_1},t_1}$ mentioned, which affects the (diagram-) line labeled by $t_1$. Finally, we replace every 0 order statistical function with the exact function and thus, we get the DIA equations (taking into count the resummation process detailed in \ref{B2}).
\section{Resummation process and expansion for order 1}
\label{B2}
To explain and clarify the resumation process, let us use the first order as an example. That is , we will only consider the diagrams with only one vertex at the most: the first and second one at the right hand side of the expansion of the correlation function \ref{ExpDiagrams}. Then we make the mentioned replacements for the exact functions; so, the final equation corresponds to \footnote{ Here we show the combinatory factor "$3\times$" as it is not explicitly written in the equation at first order}:
\begin{eqnarray}
\label{Corden1}
\hat{\mathcal{L}}^{l_1}_{\bb{p_1}, t_1}
\begin{tikzpicture}
\begin{feynman}[inline=(a)]
\vertex(b);
\vertex[right=2cm of b] (d);
\diagram*{
    (b)-- [gluon, edge label=\(t_2\hspace{15mm}t_1\)] (d),
};
\end{feynman}
\end{tikzpicture}&=&
\begin{tikzpicture}
\begin{feynman}[inline=(a)]
\vertex(b);
\vertex[right=2cm of b] (d);
\diagram*{
    (b)-- [photon, edge label=\(t_2\hspace{15mm}t_1\)] (d),
};
\end{feynman}
\end{tikzpicture}
+3\times
\begin{tikzpicture}
\begin{feynman}[inline=(b)]
\vertex(a);
\vertex[right=2cm of a] (b);
\vertex[above= of b] (c);
\diagram*{
    (a)-- [gluon, edge label=\(t_2\hspace{15mm}\)] (b),
    (b)-- [gluon, half left, edge label=\(t_1\)] (c),
    (c)-- [gluon, half left, edge label=\(\)] (b),
};
\end{feynman}
\end{tikzpicture}\nonumber\\
\end{eqnarray}
and
\begin{eqnarray}
\hat{\mathcal{L}}^{l_1}_{p_1, t_1}
\begin{tikzpicture}
\begin{feynman}[inline=(a)]
\vertex(b);
\vertex[right=2cm of b] (d);
\diagram*{
    (b)-- [photon, edge label=\(t_2\hspace{15mm}t_1\)] (d),
};
\end{feynman}
\end{tikzpicture}&=&
\delta_{l_1,l_2}\delta(t_1-t_2)
+3\times\begin{tikzpicture}
\begin{feynman}[inline=(e)]
\vertex(a);
\vertex[right=2cm of a] (b);
\vertex[above= of b] (c);
\diagram*{
    (a)-- [photon, edge label=\(t_2\hspace{15mm}\)] (b),
    (b)-- [gluon, half left, edge label=\(t_1\)] (c),
    (c)-- [gluon, half left, edge label=\(\)] (b),
};
\end{feynman}
\end{tikzpicture}
\nonumber\\
\end{eqnarray}
Thus, if we use the Green's function of the operator $\hat{\mathcal{L}}^{l_1}_{\bb{p_1}, t_1}$ which is nothing but adding a solid line as a right external leg, we can expand the recursive relation for the function $\begin{tikzpicture}
\begin{feynman}[inline=(a)]
\vertex(b);
\vertex[right=2cm of b] (d);
\diagram*{
    (b)-- [gluon, edge label=\(t_2\hspace{15mm}t_1\)] (d),
};
\end{feynman}
\end{tikzpicture}$ and eventually replace this expression in the loop and leg of the second term in the right hand side of equation \ref{Corden1}. Explicitly:
\begin{eqnarray}
&&\hat{\mathcal{L}}^{l_1}_{\bb{p_1}, t_1}
\begin{tikzpicture}
\begin{feynman}[inline=(a)]
\vertex(b);
\vertex[right=2cm of b] (d);
\diagram*{
    (b)-- [gluon, edge label=\(t_2\hspace{15mm}t_1\)] (d),
};
\end{feynman}
\end{tikzpicture}\nonumber\\
&&=
\begin{tikzpicture}
\begin{feynman}[inline=(a)]
\vertex(b);
\vertex[right=2cm of b] (d);
\diagram*{
    (b)-- [plain, edge label=\(t_2\hspace{15mm}t_1\)] (d),
};
\end{feynman}
\end{tikzpicture}
+
\begin{tikzpicture}
\begin{feynman}[inline=(b)]
\vertex(a);
\vertex[right=2cm of a] (b);
\vertex[above= of b] (c);
\diagram*{
    (a)-- [scalar, edge label=\(t_2\hspace{15mm}\)] (b),
    (b)-- [scalar, half left, edge label=\(t_1\)] (c),
    (c)-- [scalar, half left, edge label=\(\)] (b),
};
\end{feynman}
\end{tikzpicture}\nonumber\\
&&+\Bigg(\begin{tikzpicture}
\begin{feynman}[inline=(e)]
\vertex(a);
\vertex[right=2cm of a] (b);
\vertex[above= of b] (c);
\vertex[right=2cm of b] (d);
\diagram*{
    (a)-- [plain, edge label=\(t_2\hspace{10mm}\)] (b),
    (b)-- [scalar, half left, edge label=\(t\)] (c),
    (c)-- [scalar, half left, edge label=\(\)] (b),
    };
\end{feynman}
\end{tikzpicture}
\times
\begin{tikzpicture}
\begin{feynman}[inline=(e)]
\vertex(a);
\vertex[right=2cm of a] (b);
\diagram*{
    (a)-- [plain, edge label=\(\hspace{10mm}t_1\)] (b)
};
\end{feynman}
\end{tikzpicture}\Bigg)\nonumber\\
&&+3\Bigg(3
\begin{tikzpicture}
\begin{feynman}[inline=(b)]
\vertex(a);
\vertex[right=2cm of a] (b);
\vertex[above= of b] (c);
\vertex[right=2cm of b] (d);
\diagram*{
    (a)-- [scalar, edge label=\(t_2\hspace{15mm}\)] (b),
    (b)-- [scalar, half left, edge label=\(t\)] (c),
    (c)-- [scalar, half left, edge label=\(\)] (b),
    (b)-- [plain, edge label=\( \)] (d),
};
\end{feynman}
\end{tikzpicture}
\times
\begin{tikzpicture}
\begin{feynman}[inline=(b)]
\vertex(a);
\vertex[right=2cm of a] (b);
\vertex[above= of b] (c);
\diagram*{
    (b)-- [scalar, half left, edge label=\(t_1\)] (c),
    (c)-- [scalar, half left, edge label=\(\)] (b),
};
\end{feynman}
\end{tikzpicture}\Bigg)\nonumber\\
&&+\Bigg(\begin{tikzpicture}
\begin{feynman}[inline=(a.base)]
\vertex(a);
\vertex[right=2cm of a] (b);
\vertex[above= of b] (c);
\vertex[right=2cm of b] (d);
\vertex[above= of d] (e);
\diagram*{
    (a)-- [plain, edge label=\(t_2\hspace{15mm}t'\)] (b),
    (b)-- [scalar, half left, edge label=\(t'\)] (c),
    (c)-- [scalar, half left, edge label=\(\)] (b),
    (b)-- [plain, edge label=\(\hspace{18mm}t\)] (d),
    (d)-- [scalar, half left, edge label=\(t\)] (e),
    (e)-- [scalar, half left, edge label=\(\)] (d),
    };
\end{feynman}
\end{tikzpicture}
\times
\begin{tikzpicture}
\begin{feynman}[inline=(a.base)]
\vertex(a);
\vertex[right=2cm of a] (b);
\diagram*{
    (a)-- [plain,edge label=\(\hspace{18mm}t_1\)] (b),
};
\end{feynman}
\end{tikzpicture}\Bigg)\nonumber\\
&&+3\Bigg( 
\begin{tikzpicture}
\begin{feynman}[inline=(b)]
\vertex(a);
\vertex[right=2cm of a] (b);
\vertex[above= of b] (c);
\vertex[right=2cm of b] (d);
\diagram*{
    (a)-- [scalar, edge label=\(t_2\hspace{15mm}\)] (b),
};
\end{feynman}
\end{tikzpicture}
\times
3
\begin{tikzpicture}
\begin{feynman}[inline=(a)]
\vertex(b);
\vertex[right=2cm of a] (b);
\vertex[above= of b] (c);
\vertex[above= of c](d);
\vertex[right= of b] (e);
\diagram*{
    (b)-- [scalar, half left, edge label=\(t_1\)] (c),
    (c)-- [scalar, half left, edge label=\(t\)] (d),
    (d)-- [scalar, half left, edge label=\(\)] (c),
    (c)-- [plain, half left, edge label=\(\)] (b),
};
\end{feynman}
\end{tikzpicture}\Bigg)
\nonumber\\
&&+3\Bigg(3
\begin{tikzpicture}
\begin{feynman}[inline=(b)]
\vertex(a);
\vertex[right=2cm of a] (b);
\vertex[above= of b] (c);
\vertex[right=2cm of b] (d);
\diagram*{
    (a)-- [plain, edge label=\(t_2\hspace{15mm}\)] (b),
    (b)-- [scalar, half left, edge label=\(t\)] (c),
    (c)-- [scalar, half left, edge label=\(\)] (b),
    (b)-- [scalar, edge label=\( \)] (d),
};
\end{feynman}
\end{tikzpicture}
\times
\begin{tikzpicture}
\begin{feynman}[inline=(b)]
\vertex(a);
\vertex[right=2cm of a] (b);
\vertex[above= of b] (c);
\diagram*{
    (b)-- [scalar, half left, edge label=\(t_1\)] (c),
    (c)-- [scalar, half left, edge label=\(\)] (b),
};
\end{feynman}
\end{tikzpicture}\Bigg)\nonumber\\
&&+\Bigg(\begin{tikzpicture}
\begin{feynman}[inline=(a)]
\vertex(b);
\vertex[right=2cm of a] (b);
\vertex[above= of b] (c);
\vertex[above= of c](d);
\vertex[right= of b] (e);
\diagram*{
    (a)-- [plain, edge label=\(t_2\hspace{10mm}\)] (b),
    (b)-- [plain, half left, edge label=\(t'\)] (c),
    (c)-- [scalar, half left, edge label=\(t\)] (d),
    (d)-- [scalar, half left, edge label=\(\)] (c),
    (c)-- [scalar, half left, edge label=\(\)] (b),
    };
\end{feynman}
\end{tikzpicture}
\times
\begin{tikzpicture}
\begin{feynman}[inline=(a)]
\vertex(b);
\vertex[right=2cm of a] (b);
\diagram*{
    (a)-- [plain, edge label=\(\hspace{10mm}t_1\)] (b),
};
\end{feynman}
\end{tikzpicture}\Bigg)+\ldots\nonumber\\
\end{eqnarray}
where all diagrams up to order 2 are written . It should be clearly noted that the diagrams that are multiplied from the right with solid line are the ones corresponding to the expansion of the response function \footnote{These are also the ones involving an extra time variable in comparison with the diagrams of the same order}. In this way, we observe that a subset of diagrams, already included in the original expansion, appear in this last expansion. \ref{ExpDiagrams} and it is only because of the resummation for the first order term. Despite the appearing subset of diagrams, we can also observe that there are diagrams that can not be made up to this order. In consequence, the resummation replacement for \ref{ExpDiagrams} will only take place at those diagrams that cannot be recovered from the order 1
\section{Kinetic Equation and frequency correction}
\label{C}
Since the kinetic equation corresponds to the slow evolution of the \textit{same-time} correlation function, an appropriate change of variables must be made to capture the required dynamics. First of all, $(t_1,t_2)\rightarrow(\tau,t_2)$ where $\tau:=t_2-t_1$ but then as we allow different time scales $(\tau,t_2)\rightarrow(\tau,t_2,T_2,\mathcal{T}_2)$ with $T_2:=\epsilon t_2$ and  $\mathcal{T}_2:=\epsilon^2t_2$. This change yields to a respective expansion for the correlation function $C^{l_1l_2}_{\bb{p}_2}(t_1,t_2)=C^{l_1l_2}_{0,\bb{p}_2}(\tau,t_2,T_2,\mathcal{T}_2)+\epsilon C^{l_1l_2}_{1,\bb{p}_2}(\tau,t_2,T_2,\mathcal{T}_2)+\epsilon^2 C^{l_1l_2}_{2,\bb{p}_2}(\tau,t_2,T_2,\mathcal{T}_2)+\ldots$ .It is to be observed that in order to justify the expansion of the correlation function for every time, it has to remain bounded for times at any order. In this way, we will find a secular equation to ensure that the constant terms will vanish and therefore, don't grow linearly with $t_2$. The usual procedure starts with the solution for the zero order functions without forcing and dissipation:
 \begin{equation}
-\frac{\partial C^{l_1l_2}_{0,\bb{p}_2}}{\partial\tau}(\tau,t_2,T_2,\mathcal{T}_2)+il_1\omega_{\bb{p}_2}C^{l_1l_2}_{0,\bb{p}_2}(\tau,t_2,T_2,\mathcal{T}_2)=0
\end{equation}
\begin{equation}
\left(\frac{\partial}{\partial\tau}+\frac{\partial}{\partial t_2}+il_2\omega_{\bb{p}_2}\right)C^{l_1l_2}_{0,\bb{p}_2}(\tau,t_2,T_2,\mathcal{T}_2)=0
\end{equation}
Then,
 \begin{eqnarray}
 \label{C_0}
  C^{l_1l_2}_{0,\bb{p}_2}(\tau,t_2,T_2,\mathcal{T}_2)&=&C^{l_1l_2}_{0,\bb{p}_2}(0,0,T_2,\mathcal{T}_2)e^{-i(l_1+l_2)\omega_{\bb{p}_2}t_2+il_1\omega_{\bb{p}_2}\tau}\nonumber\\
\end{eqnarray}
\subsection{Order $\epsilon$: Frequency shift and non-existent energy transfer}
 The equations are:
 \begin{eqnarray}
&&\left(-\frac{\partial}{\partial\tau}+il_1\omega_{\bb{p}_2}\right)C^{l_1l_2}_{1,\bb{p}_2}(\tau,t_2,T_2)\\
&=&6\pi^2\sum_{s_1s_2s_3}\int L^{l_1s_1s_2s_3}_{-\bb{p_2-k_2k_2-p_2}}\nonumber\\
&&\times C^{s_1s_2}_{0,\bb{k}_2}(0,t_2-\tau,T_2)d\bb{k}_2C^{s_3l_2}_{0,\bb{p}_2}(\tau,t_2,T_2)\nonumber\\
\end{eqnarray}
\begin{eqnarray}
&&\left(\frac{\partial}{\partial\tau}+\frac{\partial}{\partial t_2}+il_2\omega_{\bb{p}_2}\right)C^{l_1l_2}_{1,\bb{p}_2}(\tau,t_2,T_2)\nonumber\\
&=&-\frac{\partial C^{l_1l_2}_{0,\bb{p}_2}}{\partial T_2}\nonumber\\
&&+6\pi^2\sum_{s_1s_2s_3}\int L^{l_2s_1s_2s_3}_{\bb{p_2-k_2k_2p_2}}C^{s_1s_2}_{0,\bb{k_2}}(0,t_2,T_2)C^{l_1s_3}_{0,\bb{p}_2}(\tau,t_2,T_2)\nonumber\\
\end{eqnarray}
If we add the equations we obtain:
\begin{eqnarray}
&&\left(\frac{\partial}{\partial t_2}+i(l_1+l_2)\omega_{\bb{p}_2}\right)C^{l_1l_2}_{1,\bb{p}_2}(\tau,t_2,T_2)\nonumber\\
&=&-\frac{\partial C^{l_1l_2}_{0,\bb{p}_2}}{\partial T_2}\nonumber\\
&&+6\pi^2\sum_{s_1s_2s_3}\int L^{l_2s_1s_2s_3}_{\bb{p_2-k_2k_2p_2}}C^{s_1s_2}_{0,\bb{k}_2}(0,t_2,T_2)C^{l_1s_3}_{0,\bb{p}_2}(\tau,t_2,T_2)\nonumber\\
&&+6\pi^2\sum_{s_1s_2s_3}\int L^{l_1s_1s_2s_3}_{\bb{-p_2-k_2k_2-p_2}}d\bb{k}_2\nonumber\\
&&\times C^{s_1s_2}_{0,\bb{k}_2}(0,t_2-\tau,T_2)d\bb{k}_2C^{s_3l_2}_{0,\bb{p}_2}(\tau,t_2,T_2)\nonumber\\
\end{eqnarray}
For the case $l_2=-l_1$, the only constant terms with respect to $t_2$ are: $\frac{\partial C^{l_1-l_1}_{\bb{p}_2,0}}{\partial T_2}(0,0,T_2)$ ,the first nonlinear term for the case $s_1=-s_2$,$s_3=-l_1$ and finally, the second nonlinear term for the case $s_1=-s_2$,$s_3=l_1$. Then, because of properties of the scattering coefficient $$L^{-l_1s_1s_2-l_1}_{\bb{p_2-k_1k_1p_2}}=-L^{l_1s_1s_2l_1}_{-\bb{p_2-k_1k_1-p_2}}$$ therefore, the nonlinear terms cancel each other, yielding to the secular condition:
\begin{equation}
\frac{\partial C^{l_1-l_1}_{0,\bb{p}_2}}{\partial T_2}(0,0,T_2)=0
\end{equation}
which shows that there is no energy cascade at this order. For the case $l_2=l_1$ the procedure is similar; therefore we only present the secular condition:
\begin{eqnarray}
&& \frac{\partial C^{l_1l_1}_{0,\bb{p_2}}}{\partial T_2}(0,0,T_2)\\
 &=& 6\pi \sum_{s_1}\int L^{l_1s_1-s_1l_1}_{\bb{p_2-k_2k_2p_2}}C^{s_1-s_1}_{0,\bb{k}_2}(0,0,T_2)C^{l_1l_1}_{0,\bb{p}_2}(0,0,T_2)\nonumber\\
 &&+6\pi \sum_{s_1}\int L^{l_1s_1-s_1l_1}_{\bb{-p_2-k_2k_2-p_2}}C^{s_1-s_1}_{0,\bb{k}_2}(0,0,T_2)C^{l_1l_1}_{0,\bb{p}_2}(0,0,T_2)\nonumber\\
 \end{eqnarray}
 which leads to a frequency correction of order $\epsilon$:
 \begin{equation}
C^{l_1l_1}_{0,\bb{p_2}}(0,0,T)= C^{l_1l_1}_{0,\bb{p}_2}(0,0,0)e^{i \tilde{\omega}^{l_1}_{\bb{p}_2}T}
\end{equation}
with \begin{small}$$\tilde{\omega}^{l_1}_{\bb{p}_2}\coloneqq -6\pi i\sum_{s_1}\int \big( L^{l_1l_1-s_1s_1}_{\bb{-p_2-p_2-k_1k_1}}+L^{l_1l_1-s_1s_1}_{\bb{p_2p_2-k_1k_1}}\big)C^{-s_1s_1}_{0,\bb{k}_1}(0,0,0)d\bb{k}_1$$\end{small}
\subsection{Order $\epsilon^2$: Kinetic equation}
The second order equations read as:
\begin{small}
\begin{eqnarray}
&&\big[-\frac{\partial}{\partial\tau}+il_1\omega_{p_2}\big]C^{l_1l_2}_{2,\mathbf{p}_2}(\tau,t_2,T_2,\mathcal{T}_2)\nonumber\\
&=&6\pi\sum_{s_1s_2s_3}\int L^{l_1s_1s_2s_3}_{-\mathbf{p}_2-\mathbf{k}_2\mathbf{k}_2-\mathbf{p}_2}\nonumber\\
&&\times C^{s_1s_2}_{0,\mathbf{k}_2}(0,t_2-\tau,T_2,\mathcal{T}_2)C^{s_3l_2}_{0,\mathbf{p}_2}(\tau,t_2,T_2,\mathcal{T}_2)d\mathbf{k}_{2}\nonumber\\
&&+18(2\pi)^3\sum_{s_1s_2s_3} \int d\mathbf{k}_{123}L^{l_1s_1s_2s_3}_{\mathbf{p}_1\mathbf{k}_1\mathbf{k}_2\mathbf{k}_3} \sum_x \int^{t_2-\tau}_0 dt\sum_{x_1x_2x_3} L^{xx_1x_2x_3}_{\mathbf{k}_1-\mathbf{k}_3-\mathbf{k}_2-\mathbf{p}_2}\nonumber\\
&&\times  C^{x_3l_2}_{0,\mathbf{p}_2}(t_2-t,t_2,T_2,\mathcal{T}_2)C^{x_1s_3}_{0,\mathbf{k}_3}(t_2-\tau-t,t_2-\tau,T_2,\mathcal{T}_2)\nonumber\\
&&\times C^{x_2s_2}_{0,\mathbf{k}_2}(t_2-\tau-t,t_2-\tau,T_2,\mathcal{T}_2)R^{s_1-x}_{0,\mathbf{-k}_1}(t_2-\tau-t)\delta(\mathbf{K}_{+\mathbf{p}_2})\nonumber\\
&&+6(2\pi)^3\sum_{s_1s_2s_3} \int d\mathbf{k}_{123}L^{l_1s_1s_2s_3}_{\mathbf{p}_1\mathbf{k}_1\mathbf{k}_2\mathbf{k}_3} \sum_x\int^{t_2}_0 dt\sum_{x_1x_2x_3} L^{xx_1x_2x_3}_{\mathbf{p}_2-\mathbf{k}_1-\mathbf{k}_2-\mathbf{k}_3}\nonumber\\
&&\times  C^{x_2s_2}_{0,\mathbf{k}_2}(t_2-\tau-t,t_2-\tau,T_2,\mathcal{T}_2)C^{x_1s_1}_{0,\mathbf{k}_1}(t_2-\tau-t,t_2-\tau,T_2,\mathcal{T}_2)\nonumber\\
&&\times R^{l_2-x}_{0,-\mathbf{p}_2}(t_2-t)C^{x_3s_3}_{0,\mathbf{k}_3}(t_2-\tau-t,t_2-\tau,T_2,\mathcal{T}_2)\delta(\mathbf{K}_{+\mathbf{p}_2})\nonumber\\
\end{eqnarray}
\end{small}
and
\begin{small}
\begin{eqnarray}
&&\big[\frac{\partial}{\partial\tau}+\frac{\partial}{\partial t_2}+il_2\Omega^{l_2}_{p_2}\big]C^{l_1l_2}_{2,\mathbf{p}_2}+\frac{\partial C^{l_1l_2}_{1,\mathbf{p}_2}}{\partial T_2}+\frac{\partial C^{l_1l_2}_{0,\mathbf{p}_2}}{\partial\mathcal{T}_2}\nonumber\\
&=&6\pi\sum_{s_1s_2s_3}\int L^{l_2s_1s_2s_3}_{\mathbf{p}_2-\mathbf{k}_2\mathbf{k}_2\mathbf{p}_2}\nonumber\\
&&\times C^{s_1s_2}_{0,\mathbf{k}_2}(0,t_2,T_2,\mathcal{T}_2)C^{s_3l_1}_{0,\mathbf{p}_2}(\tau,t_2,T_2,\mathcal{T}_2)d\mathbf{k}_{2}\nonumber\\
&&+18(2\pi)^3\sum_{s_1s_2s_3} \int d\mathbf{k}_{123}L^{l_2s_1s_2s_3}_{\mathbf{p}_2\mathbf{k}_1\mathbf{k}_2\mathbf{k}_3}\int^{t_2}_0 dt\sum_{xx_1x_2x_3} L^{xx_1x_2x_3}_{\mathbf{k}_1-\mathbf{k}_3-\mathbf{k}_2\mathbf{p}_2}\nonumber\\
&&\times  C^{x_3l_1}_{0,-\mathbf{p}_2}(t_2-\tau-t,t_2-\tau,T_2,\mathcal{T}_2)C^{x_1s_3}_{0,\mathbf{k}_3}(t_2-t,t_2,T_2,\mathcal{T}_2)\nonumber\\
&&\times C^{x_2s_2}_{0,\mathbf{k}_2}(t_2-t,t_2,T_2,\mathcal{T}_2)R^{s_1-x}_{0,\mathbf{-k}_1}(t_2-t)\delta(\mathbf{K}_{-\mathbf{p}_2})\nonumber\\
&&+6(2\pi)^3\sum_{s_1s_2s_3} \int d\mathbf{k}_{123}L^{l_2s_1s_2s_3}_{\mathbf{p}_2\mathbf{k}_1\mathbf{k}_2\mathbf{k}_3} \sum_x\int^{t_2-\tau}_0 dt\sum_{x_1x_2x_3} \nonumber\\
&&\times  L^{xx_1x_2x_3}_{-\mathbf{p}_2-\mathbf{k}_1-\mathbf{k}_2-\mathbf{k}_3}R^{l_1-x}_{0,\mathbf{p}_2}(t_2-\tau-t)C^{x_1s_1}_{0,\mathbf{k}_1}(t-t_2,t_2,T_2,\mathcal{T}_2)\nonumber\\
&&\times C^{x_2s_2}_{0,\mathbf{k}_2}(t-t_2,t_2,T_2,\mathcal{T}_2)C^{x_3s_3}_{0,\mathbf{k}_3}(t-t_2,t_2,T_2,\mathcal{T}_2)\delta(\mathbf{K}_{-\mathbf{p}_2})\nonumber\\
\end{eqnarray}
\end{small}
Here, we can get rid easily of derivatives with respect to $\tau$ and focus on $t_2$ for the special case where $l_2=-l_1$. We can find the secular terms by integration and then divide everything by $t_2$. Finally, we are interested in the limit of dependence of the slow scales and therefore we take $\lim\limits_{t_2 \to\infty}$:
\begin{small}
\begin{eqnarray}
&&\lim\limits_{t_2 \to\infty}\frac{1}{t_2}\int dt'_2 \bigg\{\frac{\partial}{\partial\tau}C^{l_1-l_1}_{2,\mathbf{p}_2}+\frac{\partial C^{l_1-l_1}_{1,\mathbf{p}_2}}{\partial T_2}+\frac{\partial C^{l_1-l_1}_{0,\mathbf{p}_2}}{\partial\mathcal{T}_2}\bigg\}\nonumber\\
&=&\lim\limits_{t_2 \to\infty}\int \frac{dt'_2}{t_2} \Bigg\{6\pi\sum_{s_1s_2s_3}\int L^{l_1s_1s_2s_3}_{-\mathbf{p}_2-\mathbf{k}_2\mathbf{k}_2-\mathbf{p}_2}\nonumber\\
&&\times C^{s_1s_2}_{0,\mathbf{k}_2}(0,t'_2-\tau,T_2,\mathcal{T}_2)C^{s_3-l_1}_{0,\mathbf{p}_2}(\tau,t'_2,T_2,\mathcal{T}_2)d\mathbf{k}_{2}\nonumber\\
&&+6\pi\sum_{s_1s_2s_3}\int L^{-l_1s_1s_2s_3}_{\mathbf{p}_2-\mathbf{k}_2\mathbf{k}_2\mathbf{p}_2}\nonumber\\
&&\times C^{s_1s_2}_{0,\mathbf{k}_2}(0,t_2,T_2,\mathcal{T}_2)C^{s_3l_1}_{0,\mathbf{p}_2}(\tau,t'_2,T_2,\mathcal{T}_2)d\mathbf{k}_{2}\nonumber\\
&&+18(2\pi)^3\sum_{s_1s_2s_3} \int d\mathbf{k}_{123}L^{l_1s_1s_2s_3}_{\mathbf{p}_1\mathbf{k}_1\mathbf{k}_2\mathbf{k}_3} \int^{t'_2-\tau}_0 dt\sum_{xx_1x_2x_3} L^{xx_1x_2x_3}_{\mathbf{k}_1-\mathbf{k}_3-\mathbf{k}_2-\mathbf{p}_2}\nonumber\\
&&\times  C^{x_3-l_1}_{0,\mathbf{p}_2}(t'_2-t,t'_2,T_2,\mathcal{T}_2)C^{x_1s_3}_{0,\mathbf{k}_3}(t'_2-\tau-t,t_2'-\tau,T_2,\mathcal{T}_2)\nonumber\\
&&\times C^{x_2s_2}_{0,\mathbf{k}_2}(t'_2-\tau-t,t'_2-\tau,T_2,\mathcal{T}_2)R^{s_1-x}_{0,\mathbf{-k}_1}(t'_2-\tau-t)\delta(\mathbf{K}_{+\mathbf{p}_2})\nonumber\\
&&+18(2\pi)^3\sum_{s_1s_2s_3} \int d\mathbf{k}_{123}L^{-l_1s_1s_2s_3}_{\mathbf{p}_2\mathbf{k}_1\mathbf{k}_2\mathbf{k}_3}\int^{t'_2}_0 dt\sum_{xx_1x_2x_3} L^{xx_1x_2x_3}_{\mathbf{k}_1-\mathbf{k}_3-\mathbf{k}_2\mathbf{p}_2}\nonumber\\
&&\times  C^{x_3l_1}_{0,-\mathbf{p}_2}(t'_2-\tau-t,t'_2-\tau,T_2,\mathcal{T}_2)C^{x_1s_3}_{0,\mathbf{k}_3}(t'_2-t,t'_2,T_2,\mathcal{T}_2)\nonumber\\
&&\times C^{x_2s_2}_{0,\mathbf{k}_2}(t'_2-t,t'_2,T_2,\mathcal{T}_2)R^{s_1-x}_{0,\mathbf{-k}_1}(t'_2-t)\delta(\mathbf{K}_{-\mathbf{p}_2})\nonumber\\
&&+6(2\pi)^3\sum_{s_1s_2s_3} \int d\mathbf{k}_{123}L^{l_1s_1s_2s_3}_{\mathbf{p}_1\mathbf{k}_1\mathbf{k}_2\mathbf{k}_3}\int^{t'_2}_0 dt\sum_{xx_1x_2x_3} L^{xx_1x_2x_3}_{\mathbf{p}_2-\mathbf{k}_1-\mathbf{k}_2-\mathbf{k}_3}\nonumber\\
&&\times  C^{x_2s_2}_{0,\mathbf{k}_2}(t'_2-\tau-t,t'_2-\tau,T_2,\mathcal{T}_2)C^{x_1s_1}_{0,\mathbf{k}_1}(t'_2-\tau-t,t'_2-\tau,T_2,\mathcal{T}_2)\nonumber\\
&&\times R^{-l_1-x}_{0,-\mathbf{p}_2}(t'_2-t)C^{x_3s_3}_{0,\mathbf{k}_3}(t'_2-\tau-t,t'_2-\tau,T_2,\mathcal{T}_2)\delta(\mathbf{K}_{+\mathbf{p}_2})\nonumber\\
&&+6(2\pi)^3\sum_{s_1s_2s_3} \int d\mathbf{k}_{123}L^{-l_1s_1s_2s_3}_{\mathbf{p}_2\mathbf{k}_1\mathbf{k}_2\mathbf{k}_3} \sum_x\int^{t'_2-\tau}_0 dt\sum_{x_1x_2x_3} \nonumber\\
&&\times  L^{xx_1x_2x_3}_{-\mathbf{p}_2-\mathbf{k}_1-\mathbf{k}_2-\mathbf{k}_3}R^{l_1-x}_{0,\mathbf{p}_2}(t'_2-\tau-t)C^{x_1s_1}_{0,\mathbf{k}_1}(t-t'_2,t'_2,T_2,\mathcal{T}_2)\nonumber\\
&&\times C^{x_2s_2}_{0,\mathbf{k}_2}(t-t'_2,t'_2,T_2,\mathcal{T}_2)C^{x_3s_3}_{0,\mathbf{k}_3}(t-t'_2,t'_2,T_2,\mathcal{T}_2)\delta(\mathbf{K}_{-\mathbf{p}_2})\Bigg\}\nonumber\\
\end{eqnarray}
\end{small}
The terms of order 1 will not contribute for reasons already clarified. For the second order terms, we can see that only particular cases will vanish. To see that, we have to use the Riemann-Lebesgue Lemma, because we are dealing with terms of the type: $$\lim\limits_{t \to\infty}\int^{\infty}_{-\infty}f(x)\int^t_0 e^{i\tau x}\int^{\tau}_0e^{-ix\tau'}d\tau'd\tau dx$$
Thus, we find that the secular terms will correspond to the ones where the correlation functions has opposite signs in the upper index, that is: for example for the first term of second order $x_3=l_1$, $x_1=-s_3$ and $x_2=-s_2$. On the left side of the equation we have that the only one which is resonant is $\frac{\partial C^{l_1-l_1}_{0,\mathbf{p}_2}}{\partial\mathcal{T}_2}$. Finally, the secular equation becomes:
\begin{small}
\begin{eqnarray}
\frac{\partial c^{l_1-l_1}_{\mathbf{p}_2}}{\partial \mathcal{T}_2}(\mathcal{T}_2)&=& -12\pi(2\pi)^2\sum_{s_1s_2s_3}\int d\mathbf{k}_{123}\hat{T}_{123}\left|L^{-l_1s_1s_2s_3}_{\mathbf{p}_2\mathbf{q}_3\mathbf{q}_2\mathbf{q}_1}\right|^2\nonumber\\
&&\times c^{l_1-l_1}_{\mathbf{p}_2}(\mathcal{T}_2) c^{-s_3s_3}_{\mathbf{k}_3}(\mathcal{T}_2)c^{-s_2s_2}_{\mathbf{k}_2}(\mathcal{T}_2)\nonumber\\
&&\times  \frac{s_1}{l_1}\delta(K_{-\mathbf{p}_2})\delta(l_1\omega_{p_2}+s_1\omega_{k_1}+s_2\omega_{k_2}+s_3\omega_{k_3})\nonumber\\
&&+12\pi(2\pi)^2\sum_{s_1s_2s_3}\int d\mathbf{k}_{123} \left|L^{-l_1s_1s_2s_3}_{\mathbf{p}_2\mathbf{q}_3\mathbf{q}_2\mathbf{q}_1}\right|^2\nonumber\\
&&\times c^{-s_1s_1}_{\mathbf{k}_1}(\mathcal{T}_2)c^{-s_2s_2}_{\mathbf{k}_2}(\mathcal{T}_2)c^{-s_3s_3}_{\mathbf{k}_3}(\mathcal{T}_2)\nonumber\\
&&  \delta(K_{-\mathbf{p}_2})\delta(l_1\omega_{p_2}+s_1\omega_{k_1}+s_2\omega_{k_2}+s_3\omega_{k_3})\nonumber\\
\end{eqnarray}
\end{small}
 which is the kinetic equation.
 \section{ DIA exact equations from stochastic system}
 \label{E}
We first begin with the DIA equation for the correlation function. From the nonlinear dynamic equation \ref{Stocsys} we multiply both sides by $A^{l_2}_{\bb{p}_2}(t_2)$ and ensemble average to get:
\begin{small}
\begin{eqnarray}
\mathcal{L}^{l}_{\bb{p},t}C^{ll_2}_{\bb{pp_2}}(t,t_2)&-&\sum_{s}f^{ls}_{\bb{p}}C^{sl_2}_{\bb{pp_2}}(t,t_2)-\int^{t}_{0}\sum_{x}\sigma^{lx}_{\bb{p}}(t_1,\tau)C^{xl_2}_{\bb{pp_2}}(\tau,t_2)d\tau\nonumber\\
&=&\langle\eta^{l}_{\bb{p}}(t)A^{l_2}_{\bb{p}_2}(t_2)\rangle \nonumber\\
\end{eqnarray}
\end{small}
Then, to handle the right hand side of the equation, we return to the definition of the (non-averaged) response function \cite{kraichnan1959structure}:
\begin{equation}
A^{l_2}_{\bb{p}_2}(t_2)=\sum_s\int d\bb{k}\int^{t_2}_{0}d\tau \tilde{R}^{l_2-s}_{\bb{p_2-k}}(t_2,\tau)\eta^{s}_{\bb{k}}(\tau)
\end{equation}
so we obtain for the right hand side:
\begin{equation}
\langle\eta^{l}_{\bb{p}}(t)A^{l_2}_{\bb{p}_2}(t_2)\rangle=\int^{t_2}_{0}d\tau \sum_s\int d\bb{k} R^{l_2-s}_{\bb{p_2-k}}(t_2,\tau)\langle \eta^{l}_{\bb{p}}(t)\eta^{l_2}_{\bb{p}_2}(\tau)\rangle
\end{equation}
where $\langle \tilde{R}\rangle =R$. Considering spatial homogeneity and using the imposed condition for the noise of the stochastic system, we obtain the DIA equation for the correlation function. The DIA equation for the response function can be obtained trivially by taking the funcional derivative with respect to the noise and then averaging.
 \end{multicols}
\section{Some useful identities about the response function}
\label{D}
The first identity to show is $\left\langle \frac{\delta A^{l_1}_{\bold{p_1}}(t_1)}{\delta \eta^{l_2}_{\bb{p_2}}(t_2)}\right\rangle$.
\begin{eqnarray}
\left\langle \frac{\delta A^{l_1}_{\bold{p_1}}(t_1)}{\delta \eta^{l_2}_{\bb{p_2}}(t_2)}\right\rangle&=&\int \mathcal{D}\eta P(\eta)\int \mathcal{D}[A,\tilde{A}] \frac{\delta A^{l_1}_{\bold{p_1}}(t_1)}{\delta \eta^{l_2}_{\bb{p_2}}(t_2)} e^{\sum_s\int d\bb{k}\int dt \tilde{A}^s_{\bb{k}}(t)(A^s_{\bb{k}}(t)+is\omega_{\bb{k}}A^s_{\bb{k}}(t)-\eta^s_{\bb{k}}(t)+NL) }\nonumber\\
&=& \int \mathcal{D}\eta P(\eta)\int \mathcal{D}[A,\tilde{A}]  A^{l_1}_{\bold{p_1}}(t_1) \left(-\frac{\delta e^{\sum_s\int d\bb{k}\int dt \tilde{A}^s_{\bb{k}}(t)(A^s_{\bb{k}}(t)+is\omega_{\bb{k}}A^s_{\bb{k}}(t)-\eta^s_{\bb{k}}(t)+NL)}}{\delta \eta^{l_2}_{\bb{p_2}}(t_2)}\right)\nonumber\\
&=& \int \mathcal{D}\eta P(\eta)\int \mathcal{D}[A,\tilde{A}] A^{l_1}_{\bold{p_1}}(t_1)\tilde{A}^{l_2}_{\bb{p_2}}(t_2) e^{\sum_s\int d\bb{k}\int dt \tilde{A}^s_{\bb{k}}(t)(A^s_{\bb{k}}(t)+is\omega_{\bb{k}}A^s_{\bb{k}}(t)-\eta^s_{\bb{k}}(t)+NL)}\nonumber\\
&=&\langle A^{l_1}_{\bold{p_1}}(t_1)\tilde{A}^{l_2}_{\bb{p_2}}(t_2)\rangle
\end{eqnarray}
Next, we show that the response function also corresponds to the correlation of the field with the noise:
\begin{eqnarray}
\langle A^{s'}_{\bb{k}'}(t')\eta^s_{\bb{k}}(t)\rangle &=& \int \mathcal{D}\eta P(\eta)\int \mathcal{D}[A,\tilde{A}] A^{s'}_{\bb{k}'}(t')\eta^s_{\bb{k}}(t) e^{\int -\tilde{A}(t)(\dot{A}(t)+is\Omega A(t)-\eta(t)+NL)}\nonumber\\
&=& \int \mathcal{D}\eta P(\eta)\int \mathcal{D}[A,\tilde{A}] A^{s'}_{\bb{k}'}(t')\eta^s_{\bb{k}}(t) \left.e^{\int \left[-\tilde{A}(t)(\dot{A}(t)+is\Omega A(t)-\eta(t))+b(t)\eta(t)+NL\right]}\right\rvert_{b=0}\nonumber\\
&=& \int \mathcal{D}\eta P(\eta)\int \mathcal{D}[A,\tilde{A}] A^{s'}_{\bb{k}'}(t')\frac{\delta}{\delta b^s_{\bb{k}}(t)} \left.e^{\int \left[-\tilde{A}(t)(\dot{A}(t)+is\Omega A(t)-\eta(t))+b(t)\eta(t)+NL\right]}\right\rvert_{b=0}\nonumber\\
&=& \frac{\delta}{\delta b^s_{\bb{k}}(t)}\int \mathcal{D}[A,\tilde{A}] A^{s'}_{\bb{k}'}(t')e^{\int -\tilde{A}(t)(\dot{A}(t)+is\Omega A(t)+NL)dt} \int \mathcal{D}\eta P(\eta)\left. e^{\int (s\tilde{A}(t)+b(t))\eta(t) ) dt}\right\rvert_{b=0}\nonumber\\
&=&\frac{\delta}{\delta b^s_{\bb{k}}(t)}\int \mathcal{D}[A,\tilde{A}] A^{s'}_{\bb{k}'}(t')e^{\int -\tilde{A}(t)(\dot{A}(t)+is\Omega A(t)+NL)dt} \left. e^{\int\int  (\tilde{A}(t_1)+b(t_1))2F_{\bb{k}}\delta_{s_1-s_2}\delta(\bb{k_1+k_2})\delta(t_1-t_2)(\tilde{A}(t_2)+b(t_2))}\right\rvert_{b=0}\nonumber\\
&=&\int \mathcal{D}[A,\tilde{A}] A^{s'}_{k'}(t')\left(\int  \tilde{A}^{s_1}_{\bb{k_1}}(t_1)\delta_{-s,s_1}\delta(\bb{k_1+k})\delta(t_1-t)\right)e^{\int -\tilde{A}(t)(\dot{A}(t)+is\Omega A(t)+NL)+\frac{1}{2}\int\int  \tilde{A}(t_1)\delta(t_1-t_2)\tilde{A}(t_2)}\nonumber\\
&=&\left(\int  R^{s'-s_1}_{\bb{k'-k_1}}(t',t_1)2F_{\bb{k}}\delta_{-ss_1}\delta(\bb{k+k_1})\delta(t_1-t)\right)\nonumber\\
&=& 2F_{\bb{k}}R^{s's}_{\bb{k'k}}(t',t)
\end{eqnarray}

\end{document}